\documentclass[aps, floatfix,onecolumn,preprintnumbers, superscriptaddress]{revtex4-1}
\usepackage{hyperref}
\usepackage{verbatim}
\usepackage{amsmath}
\usepackage{latexsym}
\usepackage{natbib}
\usepackage{amsmath}
\usepackage{amssymb}
\usepackage{amsthm}
\usepackage{bm}
\usepackage{color}
\usepackage{graphicx}

\usepackage{pict2e} 
\usepackage{rotating} 

\newcommand{\tr}{\mathop{\mathrm{tr}}\nolimits}

\newcommand{\ra}{\rightarrow}

\newcommand{\nn}{\nonumber}

\newcommand{\omg}{\omega}

\newcommand{\mdag}{^{\dagger}}
\newcommand{\m}[1]{\mathrm{#1}}

\newcommand{\ket}[1]{|#1\rangle}
\newcommand{\bra}[1]{\langle#1|}
\newcommand{\proj}[1]{|#1\rangle\langle#1|}

\usepackage{amsfonts}

\newcommand{\kk}{_{\bf{k}}}

\newcommand{\EE}{\mathcal{E}}

\newcommand{\px}{\sigma^x}

\newcommand{\pz}{\sigma^z}

\let\originaleqref\eqref
\renewcommand{\eqref}{Eq.~\originaleqref}

\begin{document}

\title{Breakdown of Surface Code Error Correction Due to Coupling to a Bosonic Bath}
\author{Adrian Hutter}\email{adrian.hutter@unibas.ch}
\affiliation{Department of Physics, University of Basel, Klingelbergstrasse 82, CH-4056 Basel, Switzerland}
\author{Daniel Loss}
\affiliation{Department of Physics, University of Basel, Klingelbergstrasse 82, CH-4056 Basel, Switzerland}

\date{\today}

\begin{abstract}
We consider a surface code suffering decoherence due to coupling to a bath of bosonic modes at finite temperature and study the time available before the unavoidable breakdown of error correction occurs as a function of coupling and bath parameters.
We derive an exact expression for the error rate on each individual qubit of the code, taking spatial and temporal correlations between the errors into account.
We investigate numerically how different kinds of spatial correlations between errors in the surface code affect its threshold error rate.
This allows us to derive the maximal duration of each quantum error correction period by studying when the single-qubit error rate reaches the corresponding threshold.
At the time when error correction breaks down, the error rate in the code can be dominated by the direct coupling of each qubit to the bath, by mediated subluminal interactions, or by mediated superluminal interactions.
For a 2D Ohmic bath, the time available per quantum error correction period vanishes in the thermodynamic limit of a large code size $L$ due to induced superluminal interactions, though it does so only like $1/\sqrt{\log L}$.
For all other bath types considered, this time remains finite as $L\ra\infty$.
\end{abstract}

\maketitle

\section{Introduction}

Due to its high error threshold and since it requires only nearest-neighbor gates to be performed, the surface code \cite{Dennis2002,Raussendorf2007} is the most promising platform for scalable, fault-tolerant, and universal quantum computation \cite{Fowler2012a}.
In order to test its resilience and benchmark the performance of classical algorithms for quantum error correction (QEC), the surface code is often studied with simplistic stochastic error models, where an error is an unphysical event that happens instantaneously at a specified point in space-time.
Furthermore, it is usually assumed that these errors are not spatially correlated (see, e.g., Refs.~\cite{Duclos2009,Wang2011,Fowler2012b,Wootton2012,Hutter2013,Fowler2013,Duclos2014}).
It is thus of importance to study to what degree these assumptions are satisfied for realistic models of a physical environment, and in case they are not, what the resilience of the surface code against the resulting effective error model is.

In this work, we will consider a surface code coupled to a thermal bath of freely propagating modes. A pair of recent articles \cite{Novais2013,Jouzdani2013} studied the fidelity of the surface code in this setup (at zero temperature). 
They showed that, under the assumption of a trivial error syndrome (all stabilizer operators of the code still yield a $+1$ eigenvalue), there is a sharp transition between maximal and minimal surface code fidelity as the coupling strength to the bath is increased. 
This transition provides an upper bound to the resilience of the surface code, since a logical error with a trivial error syndrome certainly cannot be corrected.

By contrast, our goal here is to find the actual time when QEC in the surface code breaks down as a function of coupling and bath parameters.
This is the time at which an error correction algorithm is no longer able to pair the surface code defects in a way that leads to a trivial operation performed on the code subspace.

In order to find these times, we follow a three-step strategy. First, calculate the error rate on each individual qubit as a function of time and physical parameters.
There are three different physical mechanisms contributing to this error rate -- the direct interaction of each qubit with the bath, subluminal interactions mediated by the bath as well as superluminal ones.
Second, study numerically how spatial correlations between such errors affect the threshold error rate of the surface code.
Third, solve for the times for which the single-qubit error rate reaches the modified threshold error rates.

When deriving actual threshold estimates, Refs.~\cite{Novais2013,Jouzdani2013} resort to the case of nearest-neighbor correlations only. 
However, we show that both subluminal and superluminal mediated long-range interactions can actually be the dominant error mechanism at the time for which the error rate reaches critical values.

%

\section{Problem and Overview}\label{sec:problem}

We consider a surface code each qubit of which is coupled to a bosonic bath at thermal equilibrium. 
In accordance with Refs.~\cite{Novais2013,Jouzdani2013}, we only consider bit-flip errors here ($\px$) and make the simplifying assumption that the bath is in thermal equilibrium at the beginning of each QEC cycle, i.e., 
that bath correlations between different QEC cycles are negligible. Physically, this can be thought of as the bath thermalizing with an even larger bath during one QEC period. 
However, we generalize the discussion in Refs.~\cite{Novais2013,Jouzdani2013} to the case of finite temperature.

Sums and  products with a tilde on top run over all surface code qubit indices $i$, while sums without a tilde are over bath modes $\textbf{k}$.
Let $H=H_0+V$ with 
\begin{align}\label{eq:H0}
H_0 = H_{\m{bos}} = \sum\kk\omg\kk a\kk\mdag a\kk,
\end{align}
and 
\begin{align}\label{eq:V}
V=\tilde{\sum}_i\px_i\otimes\frac{\lambda}{\sqrt{N}}\sum\kk|\textbf{k}|^r\left(e^{i{\bf k}{\bf R}_i}a\kk+e^{-i{\bf k}{\bf R}_i}a\kk\mdag\right),
\end{align}
where $a\kk\mdag$ ($a\kk$) are the standard creation (annihilation) operators obeying bosonic commutation relations.
Here, $\textbf{R}_i$ is the spatial location of qubit $i$ and $N=\sum\kk1$ is the number of bosonic modes of the bath. 
Physically interesting are the cases $r=0,\pm\frac{1}{2}$ \cite{Jouzdani2013}.
We consider a linear dispersion of the bath modes, $\omg\kk=v|{\bf k}|$, as is accurate for acoustic phonons, spin-waves in an antiferromagnet, or electromagnetic waves. Here, $v$ is the corresponding velocity of the modes.

Let the initial qubit density matrix be given by $\rho_q$ and the thermal state of the bath by $\rho_B\propto\exp(-\beta H_{\m{bos}})$,
where $T=1/\beta$ is the bath temperature.
The surface code requires a set of commuting many-qubit Pauli operators, called \emph{stabilizer operators}, to yield a $+1$ eigenvalue. 
All of these operators are measured at the end of each QEC cycle.
Stabilizer measurements can be performed either by applying entangling gates between code and auxiliary qubits \cite{Dennis2002,Fowler2012a} or by direct measurement of the corresponding many-qubit parity operators \cite{DiVincenzo2012,Nigg2013}.
Eigenvalues $-1$ signal that an error has occurred and are interpreted as the presence of an \emph{anyon}.
Quantum information is stored in the subspace for which all stabilizers yield a $+1$ eigenvalue.
Correspondingly, the state $\rho_q$ is restricted to this subspace, i.e., $\rho_q$ is an anyon-free state.
QEC is successful if the anyons are paired in a way which is homologically equivalent to the way they have been created.
Finding such a pairing is the task of a classical error correction algorithm \cite{Duclos2009,Wang2011,Fowler2012b,Wootton2012,Hutter2013,Fowler2013,Duclos2014}, one of which we will encounter in Sec.~\ref{sec:errcorr}.
For more details about the surface code, see Ref.~\cite{Fowler2012a}.

The decoherent evolution of the qubits is given by
\begin{align}
 \rho_q \mapsto \Phi_d(\rho_q) = \tr_B\left\lbrace e^{-iHt}(\rho_q\otimes\rho_B)e^{+iHt}\right\rbrace\ .
\end{align}
At the end of each QEC cycle, after some time $t$, we perform a measurement of all surface code stabilizer operators, which is described by the quantum channel 
\begin{align}
 \Phi_m(\sigma) = \sum_aP_a\sigma P_a\ .
\end{align}
Here, $P_a$ projects onto the space with anyon configuration $a$ and the sum runs over all possible anyon configurations $a$.

Finally, we study the state $\rho_i(t)=\tr_{\bar{i}}\circ\,\Phi_m\circ\Phi_d(\rho_q)$ of one particular qubit. Here, $\tr_{\bar{i}}$ denotes a partial trace over all qubits except qubit $i$.
Since $\rho_q$ is an anyon-free state and the stabilizer measurement projects the density matrix of the qubits to the spaces with well-defined anyon numbers, 
$\rho_i(t)$ has no contributions of terms $\px_i\rho_i$ or $\rho_i\px_i$ (here, $\rho_i=\tr_{\bar{i}}\rho_q$).
We can thus write $\rho_i(t)=(1-p_x(t))\rho_i+p_x(t)\px_i\rho_i\px_i$.

\setlength{\unitlength}{0.75mm}
\begin{picture}(60,40)
\thicklines

\put(15,20){\large $\rho_q$}

\put(27,22){decoherence}
\put(26,20){\vector(1,0){30}}
\put(37,15){$\Phi_d$}

\put(65,27){syndrome}
\put(65,22){measurement}
\put(65,20){\vector(1,0){30}}
\put(75,15){$\Phi_m$}

\put(107,27){restrict to }
\put(107,22){$i$-th qubit}
\put(104,20){\vector(1,0){30}}
\put(114,15){$\tr_{\bar{i}}$}

\put(140,20){\large $\rho_i(t)=\tr_{\bar{i}}\circ\,\Phi_m\circ\Phi_d(\rho_q)$}
\put(152.5,11){\large $=(1-p_x(t))\rho_i+p_x(t)\px_i\rho_i\px_i$}

\end{picture}

Our first goal is to calculate $p_x(t)$ as a function of the time $t$, the parameters in $H$, and the bath temperature $T=1/\beta$, which is what we carry out in Sec.~\ref{sec:sc}. 
Using the results from Sec.~\ref{sec:sc}, we calculate in Sec.~\ref{sec:entangling} the exact evolution of the density matrix of two qubits coupled to the bath and discuss the use of this bath coupling as an entangling gate.

Secondly, we discuss what implications such an error rate has for surface code error correction. 
Error correction will inevitably break down once the error rate $p_x$ on each qubit surpasses a certain critical value $p_c$.
This critical value depends on the spatial correlations between errors in the code, on the classical algorithm that is employed in order to find a pairing of the anyons, and on the probability $p_m$ with which a syndrome measurement fails.
In the symmetric case of $p_x=p_m$ and for uncorrelated errors, efficient error correction algorithms are able to perform successful error correction up to a critical value of $1.9\% - 2.9\%$ \cite{Harrington2004,Duclos2014}.
In a more involved, circuit-based modelling of syndrome extraction, critical error rates are around $1\%$ \cite{Raussendorf2007,Wang2011,Fowler2012b}.

The higher $p_m$, the lower the probability of error $p_x$ for which successful correction is possible.
Following Refs.~\cite{Novais2013,Jouzdani2013}, we consider in the following the perfect measurement case $p_m=0$ for definiteness and simplicity.
Generalization to the more realistic case of $p_m>0$ is straightforward; it merely corresponds to replacing $p_c$ (or $\tilde{p}_c$, see below) by a lower value.

If the errors on different qubits are independent from each other and stabilizer measurements are flawless ($p_m=0$), error correction inevitably breaks down if $p_x(t)>p_c=10.9\%$ \cite{Dennis2002}. 
For $p_x(t)<p_c$ the probability of an error is exponentially small in $L$, the linear size of the code, if quantum error correction is performed optimally.
The problem of performing error correction in the surface code with perfect syndrome measurements can be mapped to the classical Ising model with erroneous qubits corresponding to antiferromagnetic bonds.
The critical value $p_c$ corresponds to an order-disorder transition in this model \cite{Dennis2002}.

For uncorrelated errors, the maximal duration $\tau$ of one QEC  cycle can thus be obtained by simply inverting $p_x(\tau)=p_c$, which we exemplify for an Ohmic bath in Sec.~\ref{sec:uncorr}. 
Alternative bath types are discussed in Appendix~\ref{app:different}.
When the errors on different qubits are not independent, the breakdown of error correction will in general occur at a single-qubit error probability $\tilde{p}_c$ different from $p_c$.
If the correlations between the errors on different qubits are ignored, $\tilde{p}_c$ may be lower than $p_c$.
On the other hand, taking knowledge about such correlations properly into account can even increase $\tilde{p}_c$ beyond $p_c$.
We present an efficient algorithm that is capable of doing this for a specific kind of correlations in Appendix~\ref{app:algo}.
However, we do not know the value of $\tilde{p}_c$ for the kind of correlations between errors that arise from coupling to the bosonic bath.
Still, solving $p_x(\tau)=\tilde{p}_c$ for $\tau$ will provide us with the correct scaling of $\tau$ as a function of physical parameters like the bath temperature.

Furthermore, in Sec.~\ref{sec:errcorr} we numerically find values for $\tilde{p}_c$ for different kinds of spatial correlations between errors and provide heuristic evidence 
that the value of $\tilde{p}_c$ for the errors arising due to the bath coupling does not differ drastically from $p_c$.
We also show that for correlated two-qubit errors the surface code can, due to being a degenerate code, be used to perform error correction in regimes where the entropy in the noise exceeds the information obtained from stabilizer measurements
 -- which is in contrast to the uncorrelated case.

The resulting maximal QEC cycle times $\tau$ for the more general case of correlated errors are derived in Sec.~\ref{sec:qectime}.
The obtained expressions for $\tau$ for a variety of different parameter regimes are summarized in Sec.~\ref{sec:summary}.
We conclude in Sec.~\ref{sec:conclusions}.

\section{The single-qubit error rate $p_x(t)$}\label{sec:sc}

In this section, we calculate exactly the joint unitary dynamics of the qubits in the surface code and the modes in the bosonic bath.
From this, we derive the probability of an error on each qubit $p_x(t)$ as a function of time, taking into account all correlations with errors affecting other qubits.

We have 
\begin{align}\label{eq:Phid}
 \Phi_d(\rho_q) &= \tr_B\left\lbrace e^{-iHt}(\rho_q\otimes\rho_B)e^{+iHt}\right\rbrace \nn\\
&= \tr_B\left\lbrace e^{iH_0t}e^{-iHt}(\rho_q\otimes\rho_B)e^{+iHt}e^{-iH_0t}\right\rbrace \nn\\
&= \tr_B\left\lbrace U(t)(\rho_q\otimes\rho_B)U(t)\mdag\right\rbrace\ ,
\end{align}
where $U(t)=e^{iH_0t}e^{-iHt}=\mathcal{T}e^{-i\int_0^t\m{d}t'\,V(t')}$ denotes the evolution operator in the interaction picture.
It follows directly from the Magnus expansion (cf.\,Ref.~\cite[Appendix A]{Jouzdani2013}) 
and the fact that $\left[V(t_1),[V(t_2),V(t_3)]\right]=0$ that
\begin{align}\label{eq:Ut}
 U(t) &= \exp\left\lbrace -i\int_0^t\m{d}t_1\,V(t_1) -\frac{1}{2}\int_0^t\m{d}t_1\int_0^{t_1}\m{d}t_2\,[V(t_1),V(t_2)] \right\rbrace \nn\\
&=: \exp\left\lbrace \tilde{\sum}_i\px_i\otimes X_i(t) \right\rbrace \exp\left\lbrace -\frac{i}{2}\tilde{\sum}_{ij}J_{ij}(t)\px_i\otimes\px_j\right\rbrace\ .
\end{align} 
We have defined
\begin{align}
 X_i(t) = \frac{\lambda}{\sqrt{N}}\sum\kk\frac{|\textbf{k}|^r}{\omg\kk}\left(e^{i{\bf k}{\bf R}_i}(e^{-i\omg\kk t}-1)a\kk-e^{-i{\bf k}{\bf R}_i}(e^{i\omg\kk t}-1)a\kk\mdag\right)
\end{align}
and 
\begin{align}\label{eq:Jij}
 J_{ij}(t) &= -i\frac{\lambda^2}{N}\sum\kk|{\bf k}|^{2r}\int_0^t\m{d}t_1\int_0^{t_1}\m{d}t_2\left\lbrace e^{i{\bf k}({\bf R}_i-{\bf R}_j)}e^{-i\omg\kk(t_1-t_2)} - \m{c.c.}\right\rbrace \nn\\
&= 2\lambda^2\int\m{d}{\bf k}\,\frac{|{\bf k}|^{2r}}{\omg\kk^2}\cos\left({\bf k}({\bf R}_i-{\bf R}_j)\right)\left(\sin(\omg\kk t)-\omg\kk t\right)\ . 
\end{align}
In Appendix~\ref{sec:induced}, we provide the functions $J_{ij}(t)$ for different bath types (i.e., different combinations of spatial dimension, $D=2,3$, and bath coupling, $r=0,\pm\frac{1}{2}$).

It is straightforward to show that $[X_i(t),X_j(t)]=0$ and thus we can also write 
\begin{align}\label{eq:U}
 U(t) &= \tilde{\prod}_i\exp\left\lbrace\px_i\otimes X_i(t) \right\rbrace\tilde{\prod}_{\lbrace i,j\rbrace}\exp\left\lbrace -iJ_{ij}(t)\px_i\otimes\px_j\right\rbrace \nn\\
&= \tilde{\prod}_i\left(\cosh(X_i(t))+\px_i\otimes\sinh(X_i(t))\right)\tilde{\prod}_{\lbrace i,j\rbrace}\left(\cos(J_{ij}(t))-i\sin(J_{ij}(t))\px_i\otimes\px_i\right)\ .
\end{align}
The product $\tilde{\prod}_{\lbrace i,j\rbrace}$ is over all pairs $\lbrace i,j\rbrace$, i.e., without double-counting.

We will refer to the first factor in Eqs.~(\ref{eq:Ut}) and (\ref{eq:U}) as the \emph{decoherent} part of the evolution, and to the second part as the \emph{coherent} part.
Note that only the decoherent part of the evolution will lead to a dependence of the evolution of the code on the state of the bath (in particular its temperature). 
The coherent part is, in principle, reversible and does not lead to a transfer of quantum information from the code qubits into the bath.

Inserting \eqref{eq:U} into \eqref{eq:Phid} and expanding the products can only be done if the number of qubits coupled to the bath is small.
In Sec.~\ref{sec:entangling} we consider the case of two qubits coupled to the same bath and calculate the exact evolution of the two-qubit density matrix. 
However, if the number of qubits coupled to the bath is large, we need to follow a different route.
Note that we are only interested in whether a net-error (i.e., an odd number of $\px$-errors) occurs on qubit $i$ after application of $\Phi_d$ and $\Phi_m$.
This probability can be found with an inductive argument over $N_q$, the number of qubits in the code.

Since $\rho_q$ is a state with no anyons, the syndrome measurement $\Phi_m$ eliminates all terms in $\Phi_d(\rho_q)$ that apply a different tensor product of Pauli errors `to the left' and `to the right' of $\rho_q$.
Formally, let $\ell$ label the $2^{N_q}$ possible configurations of $\px$ errors on the code and let $\xi_{\ell}$ denote the $\ell$-th error configuration. 
Then,
\begin{align}
 \Phi_m\left(\xi_{\ell_1}\rho_q\xi_{\ell_2}\mdag\right) = \delta_{\ell_1\ell_2}\xi_{\ell_1}\rho_q\xi_{\ell_1}\mdag\ .
\end{align}
Let us call terms which have the same tensor products of Pauli operators on the left and on the right and hence survive application of $\Phi_m$ `valid' terms.

Consider first the case $N_q=1$.
Then we simply have
\begin{align}
 \Phi_d(\rho_q) = \left\langle\cosh^2\left(X_i(t)\right)\right\rangle \rho_q - \left\langle\sinh^2\left(X_i(t)\right)\right\rangle \px\rho_q\px\ .
\end{align}
(Note that $X_i(t)$ is anti-Hermitian, so $\left(\sinh\left(X_i(t)\right)\right)\mdag=-\sinh\left(X_i(t)\right)$.)
We have introduced the notation $\langle O\rangle = \tr_B\lbrace O\rho_{B}\rbrace$.
Let us thus define the single-qubit decoherence rate by $p_d(t) = -\left\langle\sinh^2\left(X_i(t)\right)\right\rangle$.

Let $p_x(t)$ denote the error probability on qubit $1$. In the case of the surface code, this is then up to boundary effects the error probability on all other qubits as well.
The error probability $p_x(t)$ is the total probabilistic weight of all valid terms that apply an odd number of errors to qubit $1$.
Let $p_x(t)_{N_q}$ denote the probability of an error on qubit $1$ if there is a total number of $N_q$ qubits in the code.
Clearly, we have $p_x(t)_1=p_d(t)$. When increasing $N_q\mapsto N_q+1$, the parity of errors on qubit $1$ is only changed if a pair of errors is applied to qubit $1$ and qubit $N_q+1$.
The weight of this happening is $\sin^2(J_{1,N_q+1}(t))$, while the weight of it not happening is  $\cos^2(J_{1,N_q+1}(t))$. This leads to the recursive formula
\begin{align}
 p_x(t)_{N_q+1} = \cos^2(J_{1,N_q+1}(t))p_x(t)_{N_q} + \sin^2(J_{1,N_q+1}(t))(1-p_x(t)_{N_q})\ .
\end{align}
Let sums and products with a prime run over all qubits except qubit $1$, i.e., from $2$ to $N_q$.
The solution is then evidently given by
\begin{align}\label{eq:pxt}
 p_x(t)_{N_q} &= \prod_i\!^{'}\cos^2(J_{1i}(t)) \times \nn\\
&\qquad
\left\lbrace
  p_d(t)\sum_{\substack{m_i\in\lbrace0,1\rbrace\\ \sum'_im_i\equiv\,0\,(\m{mod}\, 2)}} \prod_i\!^{'}(\tan^2(J_{1i}(t)))^{m_i} 
 + (1-p_d(t))\sum_{\substack{m_i\in\lbrace0,1\rbrace\\ \sum'_im_i\equiv\,1\,(\m{mod}\, 2)}} \prod_i\!^{'}(\tan^2(J_{1i}(t)))^{m_i}
\right\rbrace\ .
\end{align}

\section{Evolution of a two-qubit density matrix coupled to the bath}\label{sec:entangling}

Consider two qubits $i$ and $j$ at locations ${\bf R}_i$ and ${\bf R}_j$, respectively, that are coupled to a bosonic bath.
We assume them to be uncorrelated with the bath at $t=0$, $\rho(0)=\rho_{ij}\otimes\rho_B$.
The evolution of the two-qubit density matrix can be found using Eqs.~(\ref{eq:Phid}) and (\ref{eq:U}).
We keep the technicalities in Appendix~\ref{app:coupling} and present here the final result for the state of the two-qubit density matrix after some time $t$.
We have
\begin{align}\label{eq:2qubitfinal}
 \rho_{ij}(t) &= \left(\frac{1}{4}(1+e^{-4\Lambda(t)}\cosh(4C_{ij}(t))) +\frac{1}{2}e^{-2\Lambda(t)}\cos(2J_{ij}(t))  \right) \times \rho_{ij} \nn\\&\quad
+\left(\frac{1}{4}(1-e^{-4\Lambda(t)}\cosh(4C_{ij}(t))\right) \times (\px_i\rho_{ij}\px_i + \px_j\rho_{ij}\px_j) \nn\\&\quad
+\left(\frac{1}{4}(1+e^{-4\Lambda(t)}\cosh(4C_{ij}(t))) -\frac{1}{2}e^{-2\Lambda(t)}\cos(2J_{ij}(t))  \right) \times \px_i\px_j\rho_{ij}\px_i\px_j \nn\\&\quad
+\left(-\frac{i}{2}e^{-2\Lambda(t)}\sin(2J_{ij}(t))\right) \times \left(\px_i\px_j\rho_{ij}-\rho_{ij}\px_i\px_j\right) \nn\\&\quad
+\left(\frac{1}{4}e^{-4\Lambda(t)}\sinh(4C_{ij}(t))\right) \times \left(\px_i\px_j\rho_{ij}+\rho_{ij}\px_i\px_j-\px_i\rho_{ij}\px_j-\px_j\rho_{ij}\px_i\right)\ .
\end{align}
Here, $J_{ij}(t)$ is as defined in \eqref{eq:Jij} and
\begin{align}
 C_{ij}(t) = \langle X_i(t)X_j(t)\rangle = -\frac{\lambda^2}{N}\sum\kk|{\bf k}|^{2r}\cos\left({\bf k}({\bf R}_i-{\bf R}_j)\right)\coth(\beta\omg\kk/2)\frac{\sin^2(\omg\kk t/2)}{(\omg\kk/2)^2}\ .
\end{align}
Furthermore, we introduced the non-negative function
\begin{align}
 \Lambda(t) = - C_{ii}(t) \geq 0\ .
\end{align}
It characterizes the decoherence of each individual qubit due to its coupling to the bath and will be discussed in more detail in the next section.

Unlike the functions $J_{ij}(t)$, the functions $C_{ij}(t)$ depend on temperature. 
For $i\neq j$, they are in general hard to evaluate at finite temperature.
At zero temperature, they have been calculated for 2D baths in Ref.~\cite{Jouzdani2013}.
In the rest of this work, we will follow Ref.~\cite{Novais2013} and focus on a bath with $r=0$, $D=2$, corresponding to an Ohmic bath.
For this case, the correlator $C_{ij}(t)$ evaluates \emph{at zero temperature} to 
\begin{align}
 C_{ij}(t) = -\frac{\lambda^2}{\pi v^2} \theta(vt-R)\text{arccosh}(vt/R)\ .
\end{align}
As it turns out, however, the single-qubit error rate $p_x(t)$ depends only on the functions $\Lambda(t)$ and $J_{ij}(t)$, but not on $C_{ij}(t)$ for $i\neq j$.
For example, one easily verifies that the partial trace $\rho_i(t)$ of \eqref{eq:2qubitfinal} is independent of $C_{ij}(t)$ and, using \eqref{eq:pd} below, that the probability for a $\px$-error agrees with \eqref{eq:pxt} for $N_q=2$.
This allows us in the following sections to evaluate $p_x(t)$ without knowing the functions $C_{ij}(t)$ for $i\neq j$.

\subsection{Bath coupling as an entangling gate}
Recently, the idea of perfoming entangling gates between two qubits by coupling them to an ordered ferromagnet (which can be seen as a ``magnon bath'') and exploiting the mediated interaction has been studied in Ref.~\cite{Trifunovic2013}.
The availability of entangling gates between nearest-neighbor qubits is crucial for the circuit-based implementation of the surface code \cite{Dennis2002,Wang2011,Fowler2012a}.
Using the above result, it is straightforward to evaluate the fidelity of such a gate.
For concreteness, let us study the fidelity of maximally entangled two-qubit states (ebits) obtained using such a gate.

Consider the initial state $\rho_{ij}=\proj{0}_i\otimes\proj{0}_j$ and the maximally entangled states $\ket{\psi^\pm}=\frac{1}{\sqrt{2}}(\ket{0}_i\ket{0}_j\pm i\ket{1}_i\ket{1}_j)$.
Then,
\begin{align}
 \bra{\psi^\pm}\rho_{ij}(t)\ket{\psi^\pm} = \frac{1}{4}(1+e^{-4\Lambda(t)}\cosh(4C_{ij}(t))) \mp\frac{1}{2}e^{-2\Lambda(t)}\sin(2J_{ij}(t))\ .
\end{align}
At times for which $J_{ij}(t)$ is an odd multiple of $\pi/4$, we obtain ebits with fidelity $\frac{1}{4}(1+e^{-4\Lambda(t)}\cosh(4C_{ij}(t)))+\frac{1}{2}e^{-2\Lambda(t)}$.
For nearby qubits, $C_{ij}(t)\simeq-\Lambda(t)$, such that the fidelity simplifies to $\frac{3}{8}+\frac{1}{8}e^{-8\Lambda(t)}+\frac{1}{2}e^{-2\Lambda(t)}$.
High-fidelity ebits can thus only be obtained for times $t$ such that $\Lambda(t)\ll1$. The gate is only useful if $J_{ij}(t)$ reaches $\pi/4$ in such times.

Note that the magnon bath considered in Ref.~\cite{Trifunovic2013} has a dispersion which is parabolic rather than linear, as assumed in this work.
For a 2D Ohmic bath ($r=0$, $D=2$), the function $J_{ij}(t)$ can be calculated as described in Ref.~\cite[Appendix C]{Jouzdani2013} and evaluates to
\begin{align}\label{eq:Jijspec}
 J_{ij}(t) =
 \frac{\lambda^2}{2\pi^2v^2}\left(\theta(R-vt)\arcsin(vt/R) + \theta(vt-R)\frac{\pi}{2}\right)\ , 
\end{align}
where we have defined $R:=|{\bf R}_i-{\bf R}_j|$. 
Note that $J_{ij}(t)$ reaches a stationary value of $\frac{\lambda^2}{4\pi v^2}$ for times $t$ such that $vt>|{\bf R}_i-{\bf R}_j|$.
Choosing $\lambda=\pi v$ thus produces ebits with fidelity $\simeq1-2\Lambda(t)$ for times such that $vt>|{\bf R}_i-{\bf R}_j|$.
High-fidelity ebits are obtained in the time-interval for which $vt>|{\bf R}_i-{\bf R}_j|$ and $\Lambda(t)\ll1$, if this interval exists.

Baths in 3D behave very differently in this respect: for all values of $r=0,\pm\frac{1}{2}$, $J_{ij}(t)$ grows linearly with $t$ for $t>R/v$ in 3D (see Appendix~\ref{sec:induced}).
Similarly, $J_{ij}(t)$ grows linearly with $t$ for large enough $t$, see Sec.~\ref{sec:parabolic}.
In these cases, ebits can be obtained by maintaining the bath-coupling for a certain amount of time.

\section{Maximal QEC cycle time for uncorrelated errors}\label{sec:uncorr}

Let us now first consider the simple case where the noise on the different qubits is uncorrelated, which is relevant if the qubits are sufficiently far apart from each other such that each qubit effectively couples to its ``private bath''.
Note that for the noise to be uncorrelated, it is not enough to require that $J_{ij}(t)$ vanish for all $i$ and $j$.
The decoherent part of the evolution, too, leads to correlations between the errors on different qubits, which can be quantified by correlators $\langle X_i(t)X_j(t)\ldots X_m(t)\rangle$.
Uncorrelated noise requires that both $J_{ij}(t)\approx0$ and $C_{ij}(t)=\langle X_i(t)X_j(t)\rangle\approx0$ for all $i\neq j$.
In this case, we simply have $p_x(t)=p_d(t)$ for each qubit.

Since $X_i(t)$ is linear in the creation/annihilation operators of the bath, we can apply Wick's theorem to calculate thermal expectation values of products of the operators $X_i(t)$.
I.e., 
\begin{align}
 \left\langle X_i(t)^{2k}\right\rangle = \frac{(2k)!}{2^kk!}\left\langle X_i(t)^2\right\rangle^k\ ,
\end{align}
where $(2k-1)\times(2k-3)\times\ldots\times3\times1=\frac{(2k)!}{2^kk!}$ is the number of possible contractions.
We thus find
\begin{align}\label{eq:pd}
 p_d(t) &:= -\left\langle\sinh^2\left(X_i(t)\right)\right\rangle \nn\\
&= -\sum_{n,m=0}^\infty\frac{1}{(2n+1)!}\frac{1}{(2m+1)!}\left\langle X_i(t)^{2n+2m+2}\right\rangle \nn\\
&= -\sum_{n,m=0}^\infty\frac{1}{(2n+1)!}\frac{1}{(2m+1)!}\frac{(2n+2m+2)!}{2^{n+m+1}(n+m+1)!}\left\langle X_i(t)^2\right\rangle^{n+m+1} \nn\\
&= -\sum_{k=0}^\infty\left\langle X_i(t)^2\right\rangle^{k+1}\frac{(2k+2)!}{2^{k+1}(k+1)!}\times\underbrace{\sum_{n=0}^k\frac{1}{(2n+1)!}\frac{1}{(2k-2n+1)!}}_{2^{2k+1}/(2k+2)!} \nn\\
&= \frac{1}{2}\left(1-\exp\left\lbrace2\left\langle X_i(t)^2\right\rangle\right\rbrace\right) \nn\\
&= \frac{1}{2}\left(1-\exp\left\lbrace-2\Lambda(t)\right\rbrace\right)\ ,
\end{align}
where we have defined $k=n+m$ and $\Lambda(t)=-\left\langle X_i(t)^2\right\rangle\geq0\,$.

Different baths are characterized by their spectral density function
\begin{align}
 J(\omg) = \frac{\lambda^2}{N}\sum\kk|{\bf k}|^{2r}\delta(\omg-\omg\kk) = \alpha\omg^s\omg_0^{1-s}e^{-\omg/\omg_c}\ .
\end{align}
Here, $\alpha$ is a dimensionless bath strength, $\omg_0$ is a characteristic frequency of the bath, and $\omg_c$ is a high-frequency cut-off.
A bath with $s<1$ is called sub-Ohmic, one with $s=1$ is called Ohmic, and one with $s>1$ is called super-Ohmic.

The function $\Lambda(t)$ depends only on the spectral density function of the bath and its temperature, namely we have
\begin{align}\label{eq:integral}
 \Lambda(t) = \int_0^\infty\m{d}\omg\, J(\omg)\coth\left(\beta\omg/2\right)\frac{\sin^2\left(\omg t/2\right)}{(\omg/2)^2}\ .
\end{align}
We see that for $s\geq1$ a finite $\omg_c$ is necessary to ensure the convergence of \eqref{eq:integral}.
With a linear dispersion, $\omg\kk=v|{\bf k}|$, and a $D$-dimensional bath, we have $s=D+2r-1$.

For uncorrelated errors, surface code error correction breaks down if $p_x(t)>p_c=10.9\%$ \cite{Dennis2002}. Inverting \eqref{eq:pd}, we thus find the maximal time $\tau$ of one error correction cycle from
\begin{align}\label{eq:tauEq}
 \Lambda(\tau) = \frac{1}{2}\log\frac{1}{1-2p_c} \simeq 0.123\ .  
\end{align}
This solves the problem up to evaluation of the integral in \eqref{eq:integral} and inversion of \eqref{eq:tauEq}.


Following Ref.~\cite{Novais2013}, we restrict in the main text to the case $D=2$ and $r=0$, corresponding to an Ohmic bath.
The dimensionless bath strength parameter evaluates in this case to $\alpha=\frac{\lambda^2}{2\pi v^2}$. 
The functions $\Lambda(t)$ for the remaining combinations of $D=2,3$ and $r=0,\pm\frac{1}{2}$ are presented in Appendix~\ref{sec:decoherence}.

For the integral in \eqref{eq:integral}, we find with $s=1$ and $\beta\omg_c\gg1$, using $\coth(x)=1+2\sum_{n=1}^\infty e^{-2nx}$,
\begin{align}\label{eq:Lambda}
 \Lambda(t) &= \int_0^\infty\m{d}\omg\, \alpha\omg e^{-\omg/\omg_c}\coth\left(\beta\omg/2\right)\frac{\sin^2\left(\omg t/2\right)}{(\omg/2)^2} \nn\\
&= \int_0^\infty\m{d}\omg\, \alpha\omg e^{-\omg/\omg_c}\frac{\sin^2\left(\omg t/2\right)}{(\omg/2)^2} + 2\sum_{n=1}^\infty\int_0^\infty\m{d}\omg\, \alpha\omg e^{-\omg/\omg_c}e^{-2n\beta\omg/2}\frac{\sin^2\left(\omg t/2\right)}{(\omg/2)^2} \nn\\
&= \alpha\log\left[1+\omg_c^2t^2\right] + 2\alpha\sum_{n=1}^\infty\log\left[1+\frac{\omg_c^2t^2}{(1+n\beta\omg_c)^2}\right] \nn\\
&\simeq \alpha\log\left[1+\omg_c^2t^2\right] + 2\alpha\sum_{n=1}^\infty\log\left[1+\frac{t^2}{n^2\beta^2}\right] \nn\\
&= \alpha\log\left[1+\omg_c^2t^2\right] + 2\alpha\log\left[\frac{\beta}{\pi t}\sinh(\frac{\pi t}{\beta})\right]\ .
\end{align}
Inserting this into \eqref{eq:pd} yields
\begin{align}
 p_d(t)=\frac{1}{2}-\frac{1}{2}\left[(1+\omg_c^2t^2)\frac{\sinh^2(\pi t/\beta)}{(\pi t/\beta)^2}\right]^{-2\alpha}\ ,
\end{align}
which for non-vanishing times ($t\gg\frac{1}{\omg_c}$) is well-approximated by
\begin{align}\label{eq:pdFinal}
 p_d(t)=\frac{1}{2}-\frac{1}{2}\left[\frac{\beta\omg_c}{\pi}\sinh(\frac{\pi t}{\beta})\right]^{-4\alpha}\ .
\end{align}
Inverting $p_d(\tau)=p_c$ leads to our final solution
\begin{align}\label{eq:tauUncorr}
 \tau = \frac{\beta}{\pi}\text{arcsinh}\left[\frac{\pi}{\beta\omg_c}(1-2p_c)^{-1/4\alpha}\right]\ .
\end{align}

\section{Surface code error correction for spatially correlated errors}\label{sec:errcorr}

The form of the evolution operator derived in \eqref{eq:U} reveals that the state $\Phi_m\circ\Phi_d(\rho_q)$ contains correlations between the errors on arbitrary numbers of qubits.
The coherent part of the evolution affects each pair $\lbrace i,j\rbrace$ of qubits by a two-qubit error with probability $\sin^2(J_{ij}(t))$, while any set $\lbrace1,2,\ldots,m\rbrace$ of $m$ qubits suffers an $m$-qubit error with probability
$(-1)^{m}\left\langle\sinh^2(X_1(t))\ldots\sinh^2(X_m(t))\right\rangle$
due to the decoherent evolution. If the decoherent evolution were uncorrelated, this probability would be given by $(-1)^{m}\left\langle\sinh^2(X_1(t))\right\rangle\ldots\left\langle\sinh^2(X_m(t))\right\rangle$.
The difference between the two terms implies the presence of correlations: if a qubit suffers an error, nearby qubits have a higher chance of also being affected by an error than one would expect from the single-qubit error rate \eqref{eq:pxt} alone.

The threshold error rate of $p_c=10.9\%$ derived in Ref.~\cite{Dennis2002} applies in the case of uncorrelated errors.
The correlations mentioned above will change this value to an unknown threshold $\tilde{p}_c$.
A recent work studied the effect of clusters of errors on surface code correction when the probability of a certain cluster size is exponentially or polynomially suppressed \cite{Fowler2014}.
Thresholds were not studied in terms of the single-qubit error rate $p_x$ but in terms of an over-all probability $p$ for single-qubit errors and clusters of errors.
If the probability of a large cluster decays sufficiently slowly, any $p>0$ will lead to $p_x\ra\frac{1}{2}$ for large enough $L$.
This makes a direct application of the results of Ref.~\cite{Fowler2014} to our problem impossible.

In the following, we thus want to investigate how different kinds of spatial correlations between errors affect the threshold error rate for the single-qubit error rate $p_x$.
The modified threshold error rate $\tilde{p}_c$ strongly depends on the type of correlations that are present between the errors.

\begin{figure}
  \setlength{\unitlength}{0.8\textwidth}
  \begin{picture}(0.65,0.5)
	\thicklines
	\put(-0.15,0.32){\includegraphics[width=0.15\textwidth]{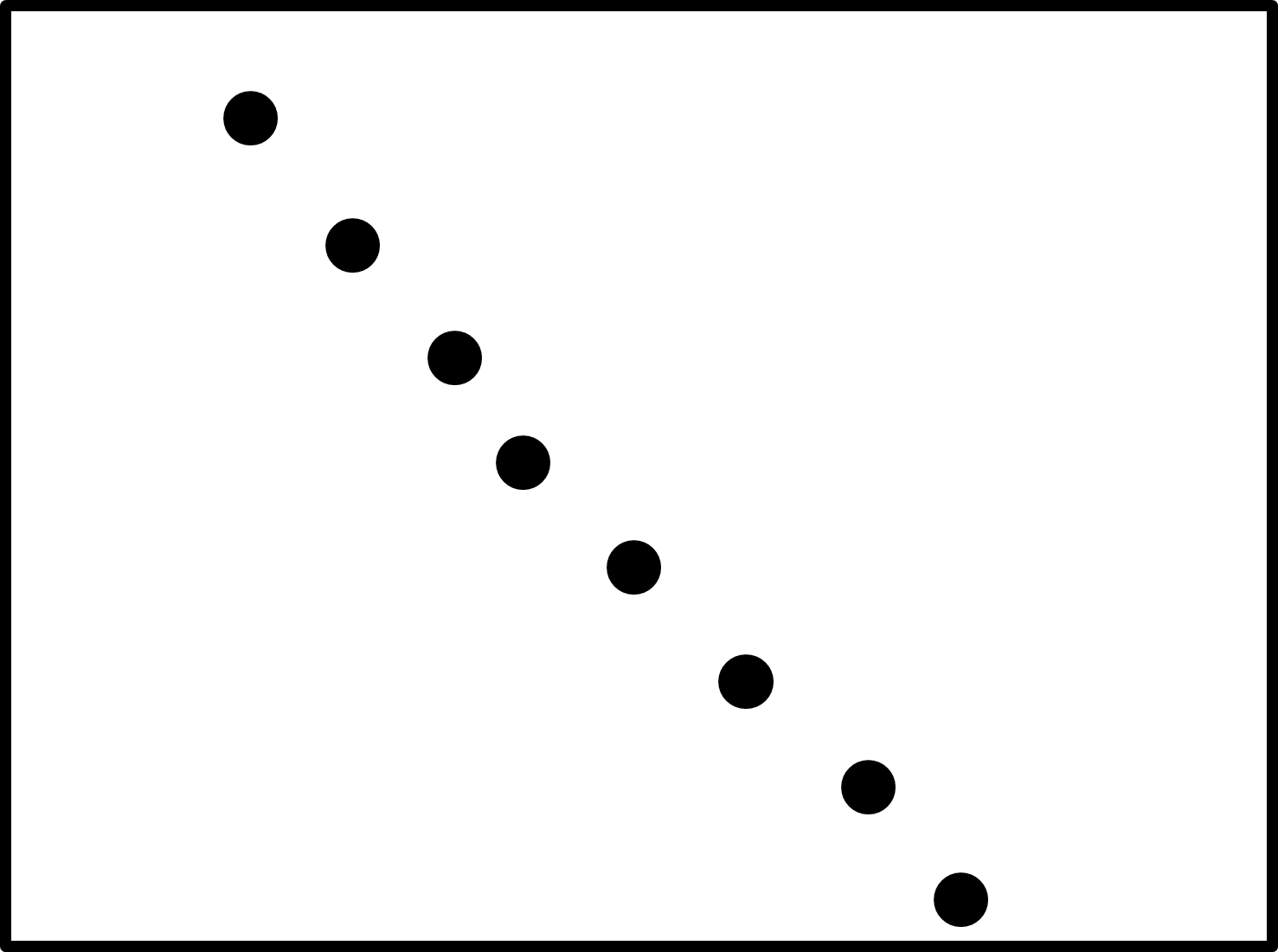}}
	\put(0.15,0.32){\includegraphics[width=0.15\textwidth]{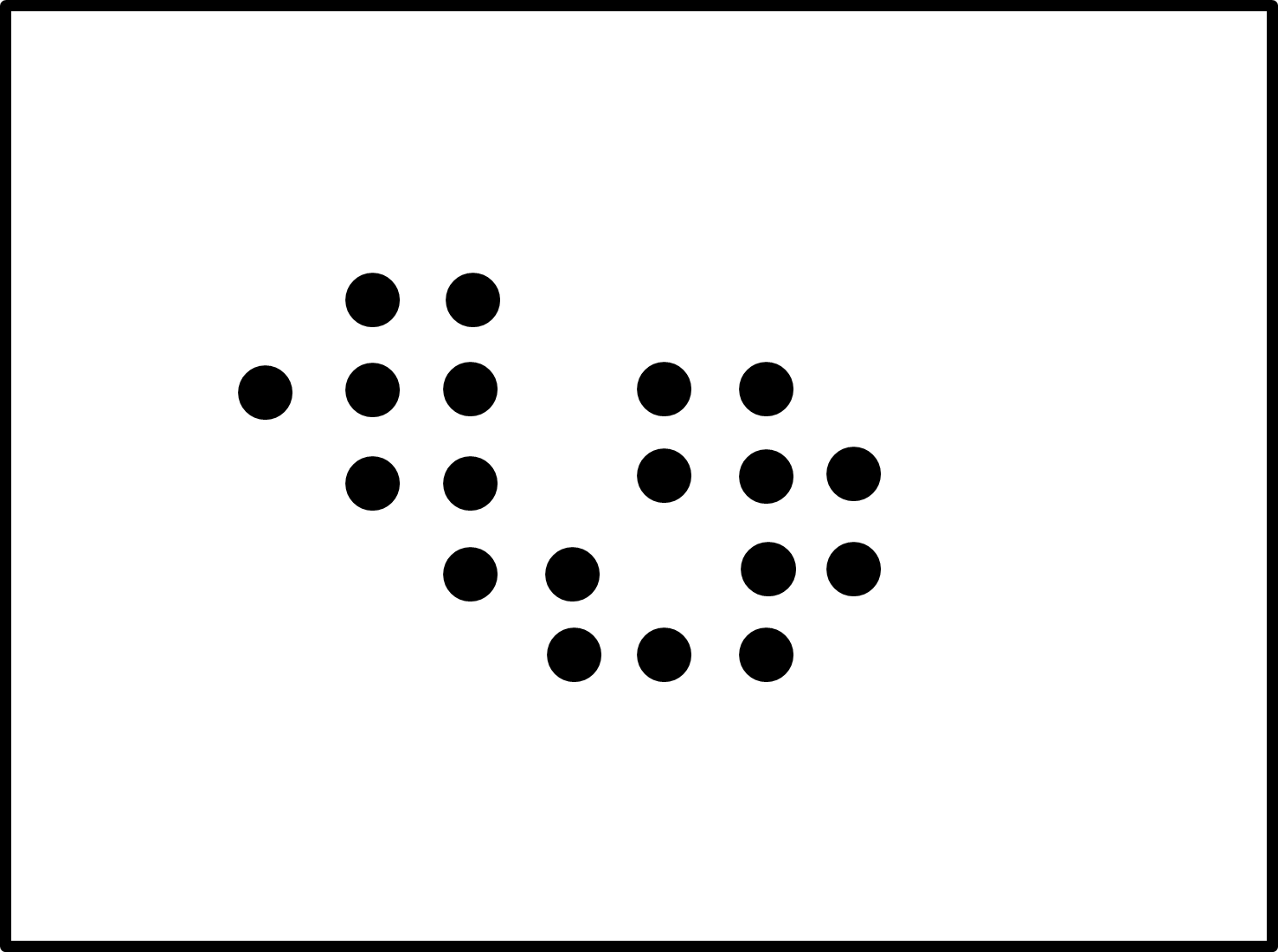}}
	\put(0.45,0.32){\includegraphics[width=0.15\textwidth]{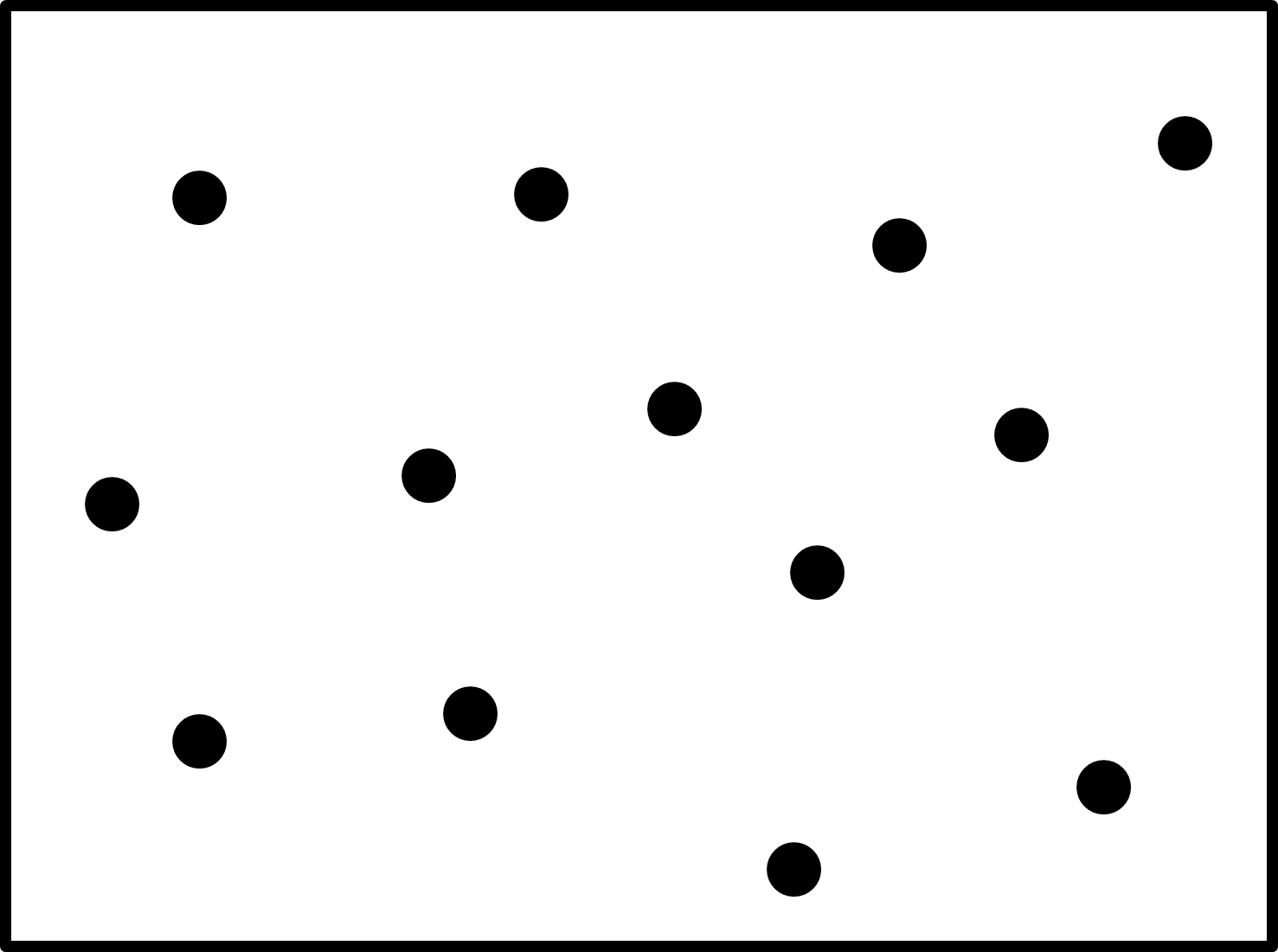}}
	\put(0.45,0.0){\includegraphics[width=0.15\textwidth]{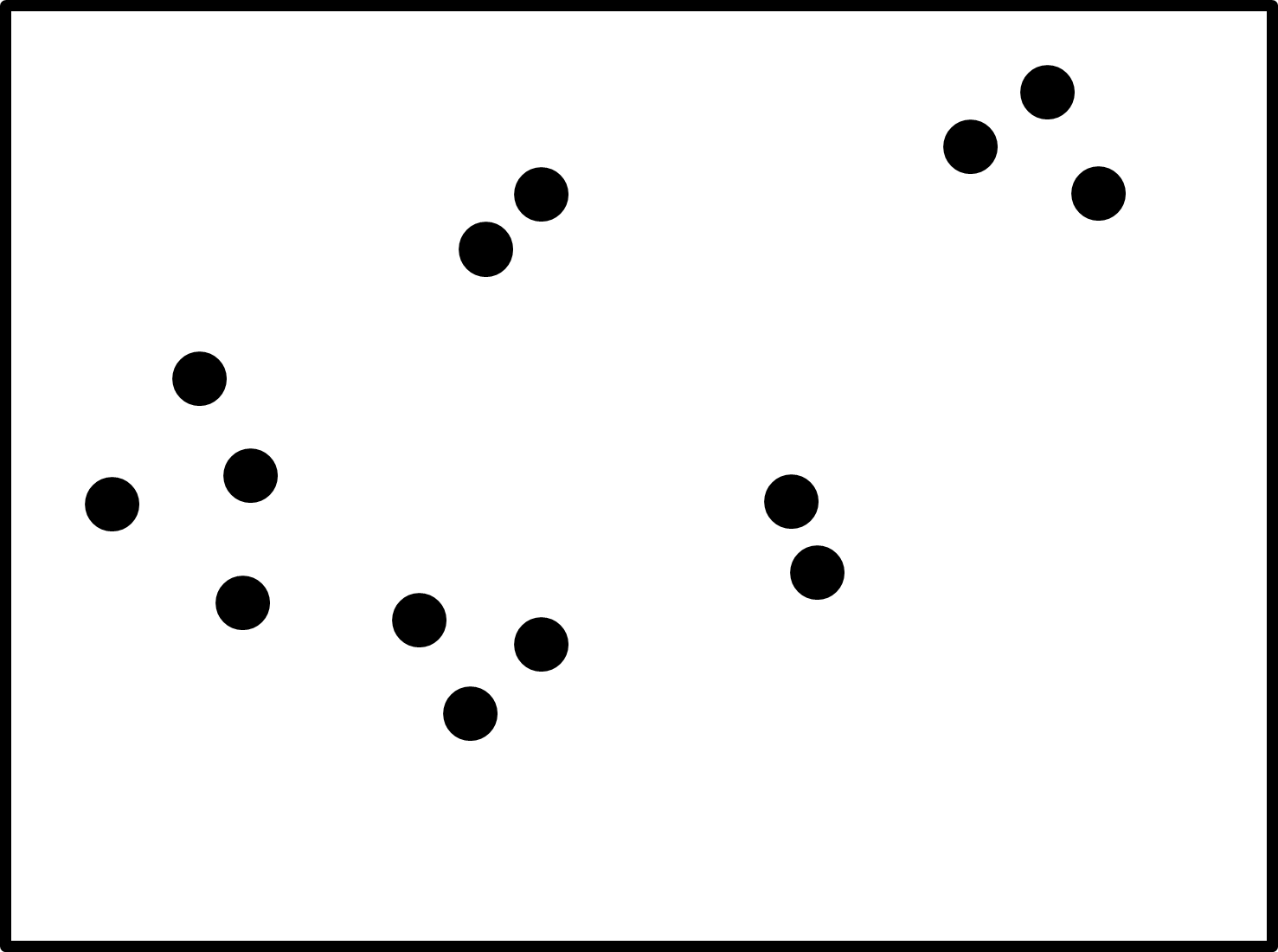}}
	\put(-0.20,0.24){\vector(1,0){0.9}}
	\put(-0.09,0.28){ballistic}
	\put(0.21,0.28){diffusive}
	\put(0.49,0.28){uncorrelated}
	\put(0.50,0.17){clustered}
	\put(-0.05,0.22){\line(0,1){0.04}}
	\put(0.25,0.22){\line(0,1){0.04}}
	\put(0.55,0.22){\line(0,1){0.04}}
	\put(0.465,0.21){\large $\underbrace{\qquad\qquad\qquad}$}
	\put(0.73,0.245){threshold}
	\put(0.73,0.220){error rate}
  \end{picture}
  \caption{Different kinds of spatial correlations between errors in the surface code and how they affect its threshold error rate.}
  \label{fig:errorCorrelations}
\end{figure}
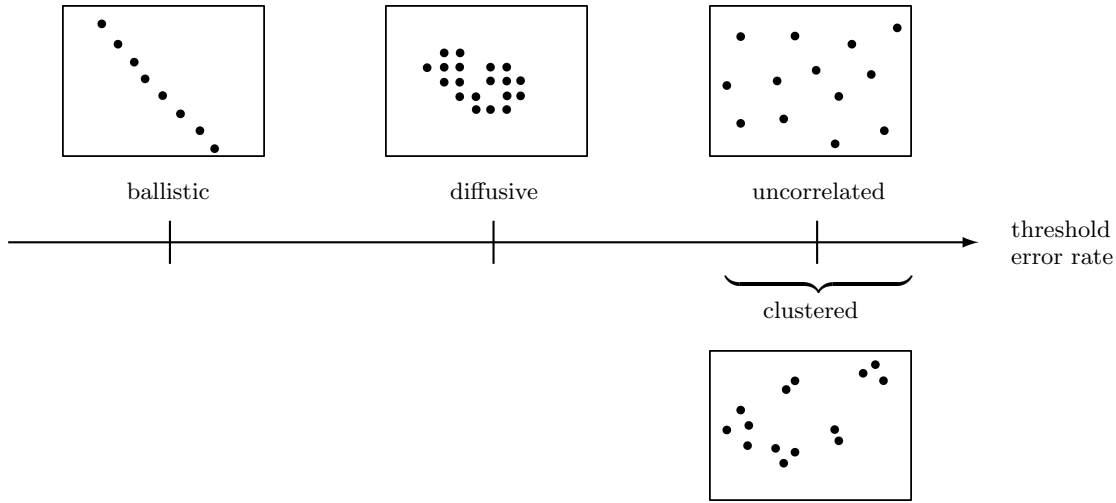

Fig.~\ref{fig:errorCorrelations} summarizes our results. A worst case is given by ballistically propagating anyons, leaving a linear trail of errors behind. In this case, $\tilde{p}_c$ can be smaller than $p_c$ by an order of magnitude or more.
To understand this, note that the task of error correction is to pair the anyons in a way that is homologically equivalent to the way they have been created.
Error correction breaks down if choosing the right homology class becomes ambiguous. This is achieved with the smallest number of errors if the anyons in each pair propagate into opposite directions.

If anyons perform a diffusive random walk in the toric code, the modified threshold error rate $\tilde{p}_c$ can also be significantly smaller than $p_c$.
This scenario is physically relevant if there is a non-trivial surface (or toric) code Hamiltonian that energetically penalizes the creation, but not the propagation of anyons.
The error model of diffusive errors and its effect on error correction have been studied in this context in Refs.~\cite{Chesi2010,Hutter2012}.

For both ballistic propagation and a diffusive random walk of anyons, there is a tendency for errors to form string-like patterns.
By contrast, the correlations discussed at the beginning of this section favor a clustering of errors (i.e., it is more likely than in the uncorrelated case that errors are spatially close to each other) but there is no mechanism that favors string-like error configurations.

We do not expect clustering of errors to strongly harm the threshold error rate $p_c$.
Most clusters of nearby errors do not form string-like patterns and thus do not help to bring pairs of anyons apart from each other and make a homologically correct pairing ambiguous.
For a fixed single-qubit error rate $p_x$, the presence of regions with a high density of errors implies the presence of regions with a low density of errors. The latter help to avoid ambiguities.

In the following subsections we study the modified threshold error rate $\tilde{p}_c$ for different kinds of spatial correlations between surface code errors by use of Monte Carlo simulations.
In agreement with our expectations, we find that clustering of errors leads to at most a mild decrease of the threshold error rate -- and can even be beneficial in the strongly correlated regime.

We conclude that even in the presence of spatial correlations between errors \emph{without a mechanism that prefers string-like arrangements} the modified threshold error rate $\tilde{p}_c$ does not differ drastically from $p_c$.
Heuristically, we expect correlations between errors arising from coupling the code to the bath not to be of the string-like type.
We will thus in the following section invert the equation $p_x(\tau)=\tilde{p}_c$ without knowing the exact value of $\tilde{p}_c$, and simply assume that it is of the same order of magnitude as $p_c$.

\subsection{Ballistic propagation of anyons}\label{sec:ballistic}

In the following subsections, we study the impact of correlated errors on the correctability of the surface code by use of Monte Carlo simulations.
That is, we produce a large number of error configurations using a certain error model, and see whether we are able to find a pairing of the resulting anyon configuration that is homologically equivalent to the actual one.
Finding such a paring is the task of a classical decoding algorithm. Only if unrealistic computing power is available can we hope to actually perform correction up to the theoretical threshold of $p_c=10.9\%$ (in the uncorrelated case).
Therefore, an efficient approximate error correction algorithm is needed in practice.
We will employ minimum-weight perfect matching (MWPM) \cite{Edmonds1965}, which, for a graph with weighted edges and an even number of vertices provides the matching of minimal weight.
Here, the vertices correspond to the anyons found as a result of the stabilizer measurements, 
and the weight of an edge connecting two anyons is simply given by the minimal number of qubits that have to suffer an error in order to create that pair from the anyonic vacuum (i.e., their Manhattan distance).
We employ the library \texttt{Blossom V} \cite{Kolmogorov2009} to perform MWPM.
Using MWPM for performing error correction in the surface code reduces the threshold error rate to $10.2\%$ \cite{Chesi2010,Fowler2012b}.

For our first ``worst case'' error model, we envision anyons that after creation start to ballistically propagate into a certain direction.
More precisely, we specify the error model by two parameters $f$ and $l$.
First, we draw a number $n$ at random from a Poisson distribution with mean $2fL^2$.
Then, we perform $n$ times the following. Choose one of the $L^2$ anyon locations and an angle $\phi\in\left[0 , 2\pi\right)$ at random. 
(Recall that we consider one type of error only, so for a surface code of linear size $L$ with periodic boundary conditions, there are $L^2$ anyon locations of the relevant type.)
Draw random numbers $l_h$ and $l_v$ from Poisson distributions with mean $l|\cos(\phi)|$ and $l|\sin(\phi)|$, respectively (the expectation value for $l_h+l_v$ is thus $\frac{4}{\pi}l$).
Starting from the initial anyon location, apply $l_h$ errors horizontally and $l_v$ errors vertically, with the directions  given by the sign of the trigonometric functions.
After doing this $n$ times, perform error correction by means of MWPM.

For each value of $l$, there is a threshold value $f_c$ such that for $f<f_c$ the logical error rate decreases exponentially with $L$ and for $f>f_c$ the logical error rate approaches $\frac{1}{2}$.
For each triple of $l$, $f$, and $L$, we generate a number $N$ of error configurations which is such that error correction fails $10^4$ times. The logical error rate can then be estimated as $10^4/N$.
The threshold values $f_c$ are then determined for each value of $l$ by comparing the logical error rates for code sizes up to $L=60$.
Finally, once we know the threshold value $f_c$ we can determine the threshold $\tilde{p}_c$ for the single-qubit error rate $p_x$ by determining the fraction of qubits that suffer an error for the given pair of $l$ and $f_c$.
An even number of errors on the same qubit count as no error, and on odd number as one. 
If the errors are sufficiently sparse such that the probability of several errors happening on the same qubit is negligible, we have $p_x=\frac{4}{\pi}l\times 2fL^2/(2L^2)=\frac{4}{\pi}lf$, while otherwise it will be smaller.

\begin{figure}[h]
	\centering
		\includegraphics[width=0.6\textwidth]{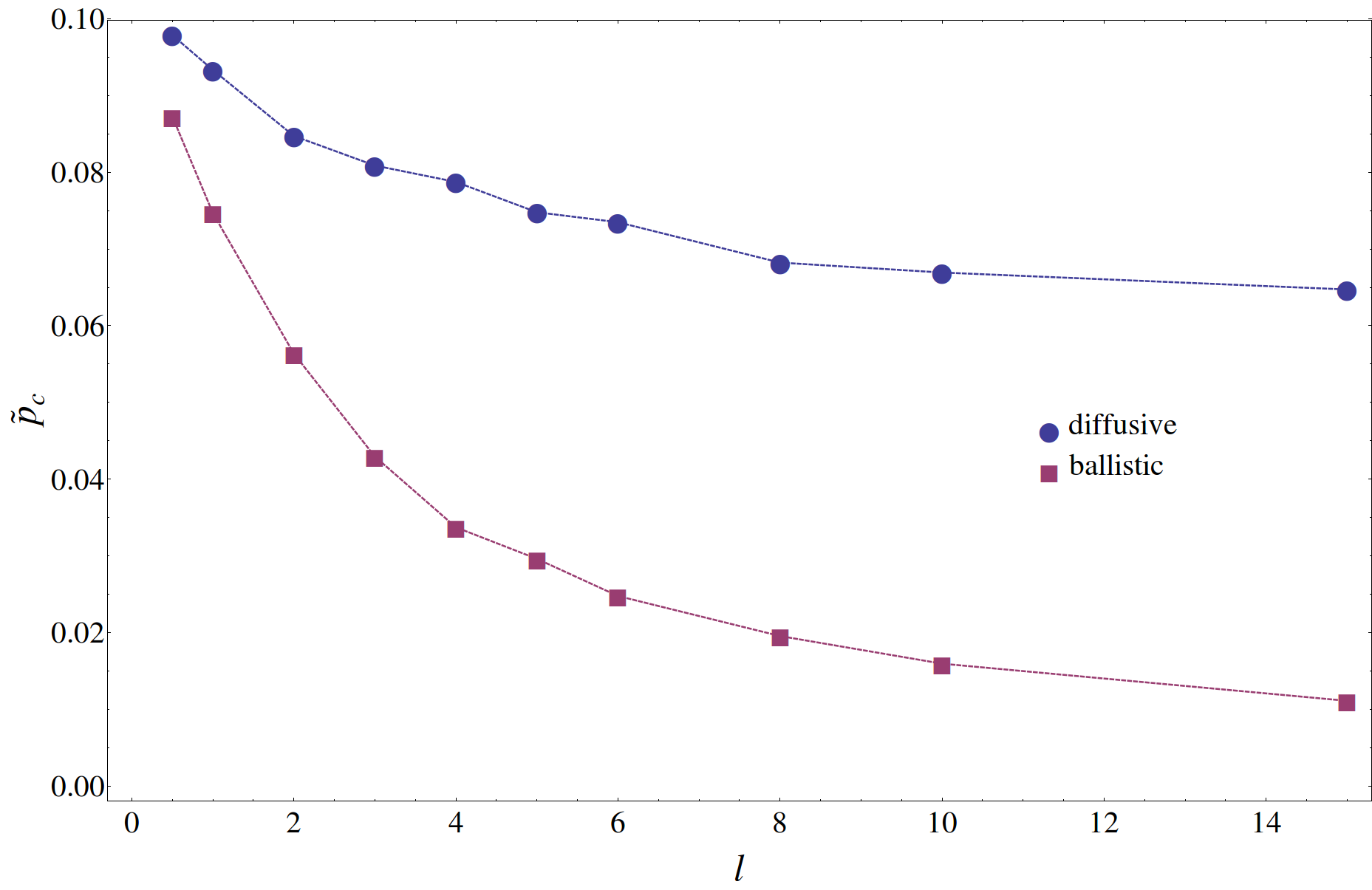}
	\caption{Single-qubit error rate $\tilde{p}_c$ for which error correction breaks down for two error models that lead to string-like error patterns: ballistic and diffusive propagation of anyons.}
	\label{fig:stringlike}
\end{figure}

The single-qubit threshold error rates $\tilde{p}_c$ as a function of $l$ are illustrated by the purple squares in Fig.~\ref{fig:stringlike}.
While for $l=\frac{1}{2}$ the threshold is still comparable with the value of $10.2\%$ for the uncorrelated case, it decreases strongly as $l$ is increased.

\subsection{Diffusive propagation of anyons}

In the case where anyons perform a random walk, the simulation works in much the same way as described in the previous subsection.
For each initial anyon location, we draw a random number from a Poisson distribution with mean $l$, and then perform a random walk whose length is given by this number.
The resulting thresholds are displayed by the blue circles in Fig.~\ref{fig:stringlike}. Threshold error values are, for a given value of $l$, significantly higher than in the ballistic case though significantly lower than in the uncorrelated case.

\subsection{Clustered errors}

Here, we study a family of error models that describe clustering of errors in the surface code.
For $l\leq m^2$, we define the error model $m$-$l$-cluster as follows: from each square of $m\times m$ qubits in the surface code, pick $l$ qubits at random and apply an error to all of them with probability $f$.
The resulting single-qubit error rate is $p_x\lesssim fl$. (Note that the same qubit can suffer several errors and an even number corresponds to no error at all, leading to $p_x<fl$.)
The modified critical error rates $\tilde{p}_c$ are again determined as described in Sec.~\ref{sec:ballistic}.

\begin{figure}[h]
	\centering
		\includegraphics[width=0.6\textwidth]{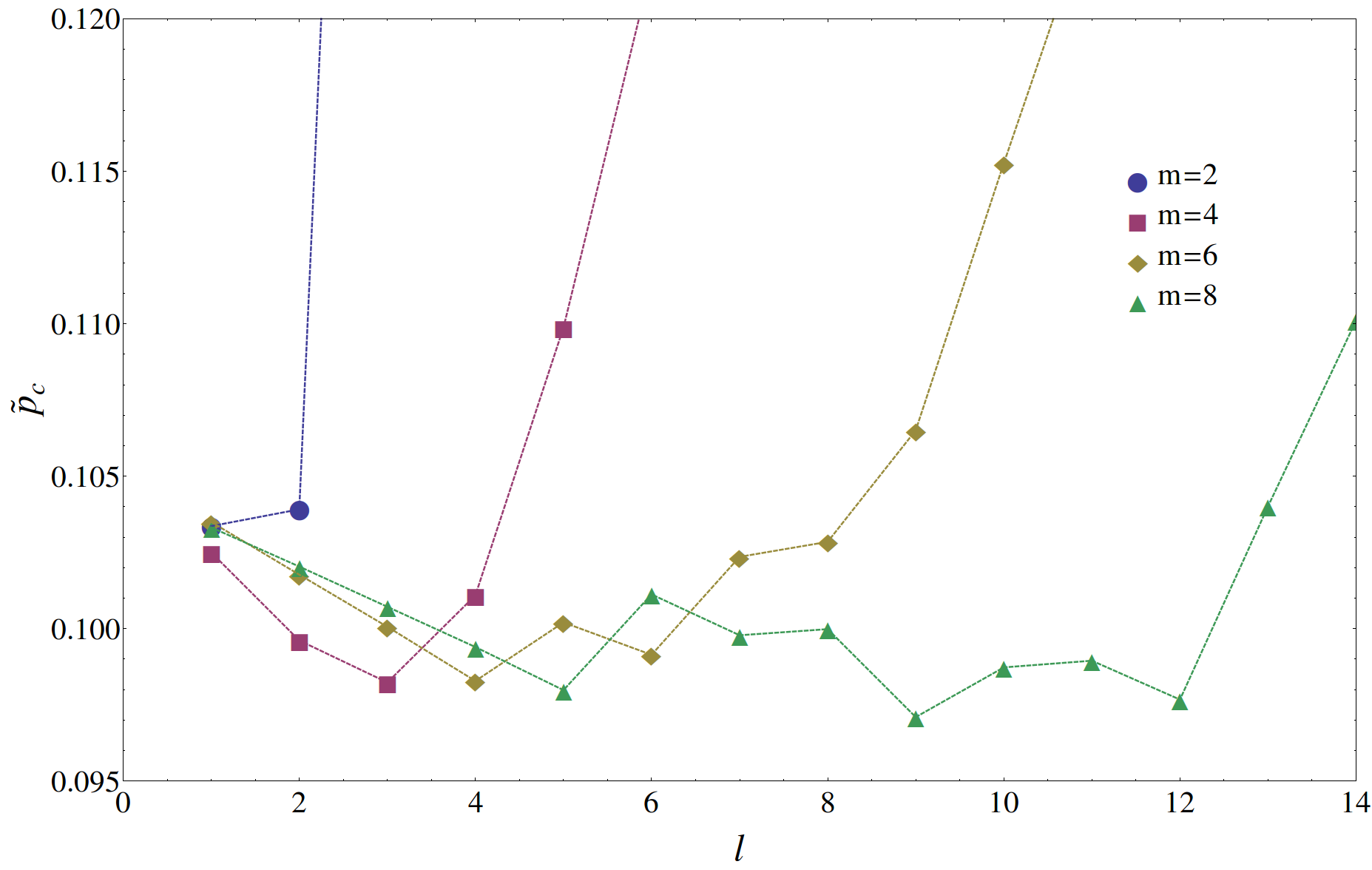}
	\caption{Single-qubit error rate $\tilde{p}_c$ for which error correction breaks down in the $m$-$l$-cluster error models.}
	\label{fig:cluster}
\end{figure}

Fig.~\ref{fig:cluster} shows our results.
If $l\ll m$, errors are essentially uncorrelated and the threshold values for $\tilde{p}_c$ are close to $10.2\%$, the threshold for MWPM-based error correction in the uncorrelated case.
For $l\lesssim m$, $\tilde{p}_c$ falls slightly below $10\%$, though the decrease is not dramatic.
This decrease is due to the possibility of forming string-like patterns of length $l$, which leads to a smaller number of errors being necessary for correction to become ambiguous.
Finally, for $l>m$ the threshold increases significantly beyond $p_c$. Additional errors now make it easier to recognize the cluster and increase the probability that several errors together form a (partial) stabilizer operator and therefore do no harm to the code.
For instance, in the $2$-$4$-cluster case the threshold error rate is as high as $\tilde{p}_c=29.0\%$, since half of all errors combine to a stabilizer operator.
In reality, we do of course not expect the environment to apply exclusively $2\times2$ squares of errors, but to find ourselves in the regime where the clustering of errors leads to a slight reduction of the single-qubit threshold error rate.

\subsection{Correlated two-qubit errors}

Let us now study the case where there are correlations between errors on pairs of qubits only.
Note that the coherent part of the evolution is able to produce such correlations only.
The regime considered here is thus relevant if correlations between error events on more than two qubits due to the decoherent evolution are weak.

The study of correlated two-qubit errors is simplified by the fact that there is a clear worst-case, namely a two-qubit error on a pair of nearest-neighbor qubits.
We assume that each qubit in the code suffers an error with probability $p_1$ and that, furthermore, each pair of nearest neighbors in the code suffers a pair of errors with probability $p_2$.
We expect and have verified in numerical simulations (see below) that correlated errors on pairs of qubits which are not nearest neighbors have, for a fixed single-qubit error rate $p_x$, less of an effect on error correction than correlated errors on nearest neighbor qubits.
Studying this particular case thus allows us to find the maximal impact of correlated two-qubit error events.

With the above parameters, and since each qubit in the code has four nearest neighbors, the single-qubit error rate $p_x$ can be calculated in analogy to \eqref{eq:pxt} as
\begin{align}\label{eq:px}
 p_x &= p_1\sum_{k\text{ even}}\binom{4}{k}p_2^k(1-p_2)^{4-k} + (1-p_1)\sum_{k\text{ odd}}\binom{4}{k}p_2^k(1-p_2)^{4-k}\nn\\
&= \frac{1}{2}-\frac{1}{2}(1-2p_1)(1-2p_2)^4\ .
\end{align}

We can make two estimates for where error correction will break down in the above model.
First, we can simply assume that the correlations do neither help nor derogate the correctability of the code. In this case, the breakdown occurs for $p_x = p_c$ (or, with MWPM correction, for $p_x=10.2\%$), independently of $p_2$.
A second estimate is of entropic nature. It is obtained by studying whether it is at all possible that the stabilizer measurements provide us with enough information to infer what errors have happened.
Assume that there are $n$ qubits in the code. There are $2n$ pairs of nearest neighbors and $n/2$ plaquette stabilizers that can give us information about bit-flip errors.
For large $n$, the total information contained in the noise can be compressed to $nh(p_1)+2nh(p_2)$ bits, where $h(p)=-p\log_2(p)-(1-p)\log_2(1-p)$ is the binary entropy function. On the other hand, the plaquette stabilizers give us at most $n/2$ bits of information.
Error correction will thus break down if
\begin{align}\label{eq:entropicEstimate}
 2h(p_1)+4h(p_2)=1\ .
\end{align}
If we needed to know exactly which qubits have suffered a bit-flip, \eqref{eq:entropicEstimate} would put a rigorous upper bound on the correctability of the surface code.
However, we only need to know the error pattern \emph{modulo application of stabilizer operators}. For this reason, \eqref{eq:entropicEstimate} should rather be seen as an estimate of an upper bound.
Such an entropic estimate predicts the unavoidable breakdown of surface code error correction to high accuracy for both uncorrelated bit-flip errors \cite{Dennis2002} (i.e., $2h(p_c)\simeq1$) and depolarizing noise \cite{Bombin2012}.
Ref.~\cite{Roethlisberger2012} shows that variations of the surface code tailored for stability against biased noise ($p_x\neq p_z$) give thresholds that fall only a few percents short of the ones suggested by such entropic arguments
 -- even with error correction performed by an efficient approximate algorithm.

We will use two different algorithms for performing error correction for the above error model.
Both of them are based on MWPM, but they differ in the weights they assign to the edges.
The first one is the algorithm used in the previous subsections. It ignores correlations and assigns the Manhattan distance between two anyons to the edge connecting them.
The second algorithm, described in more detail in Appendix~\ref{app:algo}, uses a more sophisticated assignment of edge weights that allows it to take spatial correlations between the errors into account.

\begin{figure}[h]
	\centering
		\includegraphics[width=0.8\textwidth]{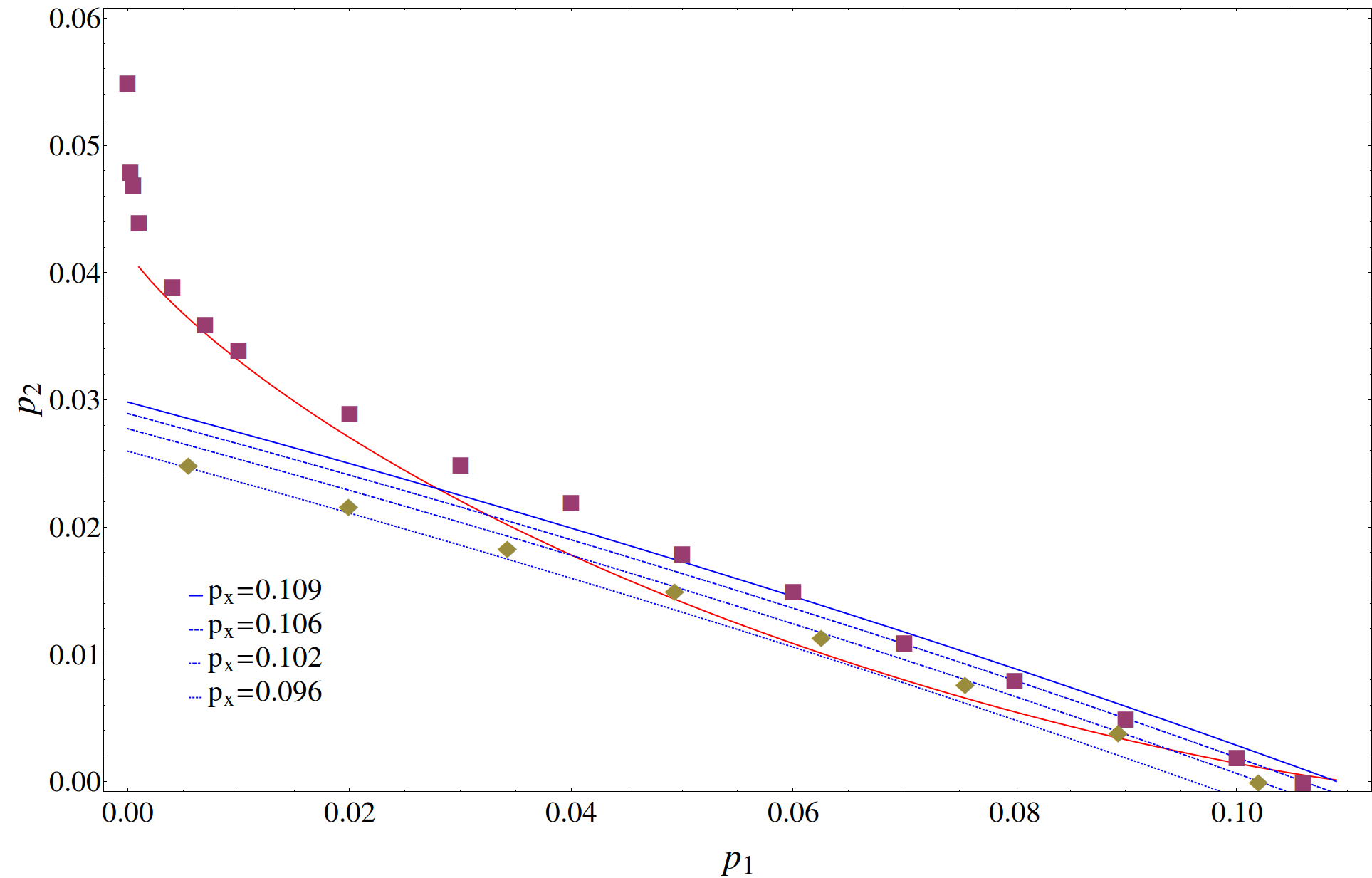}
	\caption{Each qubit is independently subjected to an error with probability $p_1$. Furthermore, each pair of nearest-neighbor qubits is subjected to a a pair of errors with probability $p_2$.
  The blue lines correspond to a constant value of $p_x$, calculated according to \eqref{eq:px}, while the red line shows the entropic bound \eqref{eq:entropicEstimate}.
  Diamonds represent threshold error rates $(p_1,p_2)$ when error correction is performed with MWPM and correlations are ignored.
  Squares represent threshold error rates for an algorithm that takes correlations into account.
  Threshold error rates have been determined to accuracy $10^{-3}$, by comparing logical error rates for code sizes between $10$ and $50$ (periodic boundary conditions).
  For each combination of error rates and code sizes, the logical error rates were obtained from as many error configurations as were necessary to obtain $10^4$ logical errors.}
	\label{fig:breakdown}
\end{figure}

Fig.~\ref{fig:breakdown} compares the above estimates with the resulting combinations $(p_1,p_2)$ for which error correction breaks down in actual numerical simulations, when the two algorithms described above are used for performing error correction.
If the Manhattan distance between two anyons is used as the edge weight and correlations between the errors are ignored, error correction breaks down for $p_x=10.2\%$ for $p_2\ra0$, slightly below the value of $p_c$ for perfect error correction.
In the maximally correlated regime, $p_1\ra0$, error correction already breaks down for $p_x=9.6\%$ -- a pretty insignificant decrease.
We have obtained similar data to the one displayed in Fig.~\ref{fig:breakdown} for correlated errors that happen on pairs of qubits which are further away from each other than nearest neighbors.
In this case, the deviations from the line $p_x=10.2\%$ are smaller.
Already for pairs of qubits that are three lattice constants away from each other, the obtained threshold error rates are indistinguishable (to accuracy $10^{-3}$) from this line.

For the second, improved algorithm, error correction breaks down for $p_x=10.6\%$ in the uncorrelated case ($p_2\ra0$), close to the theoretical value of $p_c$, and for $p_x=18.6\%$ in the maximally correlated case ($p_1\ra0$). 
The threshold error rates $(p_1,p_2)$ approximately follow that of the two above estimates wich predicts the higher threshold value and significantly beat both estimates in some regimes.
Beyond the red line in Fig.~\ref{fig:breakdown}, it is information-theoretically impossible that we learn from the stabilizer measurements what errors have happened.
That it is possible to error correct beyond that line shows that due to its degenerate nature (i.e., different error configurations can lead to the same syndrome) the surface code is able to take care of some of the entropy in the noise itself.

In conclusion, ignoring during error correction that pairs of qubits can be affected by correlated errors hardly affects the single-qubit threshold error rate of the surface code.
If an algorithm takes these correlations into account, the single-qubit threshold error rate can be significantly boosted in the strongly correlated regime.
Due to its degenerate nature, the surface code is able to correct in regimes where it is information-theoretically impossible that we learn what errors the code has suffered.

\section{Maximal QEC cycle time for correlated errors}\label{sec:qectime}

Assuming that the form of spatial correlations between errors that will be present in $\Phi_m\circ\Phi_d(\rho_q)$ does not lead to a threshold error rate $\tilde{p}_c$ that differs drastically from $p_c$,
the single-qubit error rate $p_x(t)$ in \eqref{eq:pxt} contains already all the information we need in order to predict the maximal QEC period $\tau$.
A great advantage of \eqref{eq:pxt} is that it depends only on $p_d(t)$ and the coherent interaction strengths $J_{ij}(t)$, but not on the temperature-dependent correlators $C_{ij}(t)$ for $i\neq j$.

Our goal is thus to solve the equation $p_x(\tau)=\tilde{p}_c$ for $\tau$, where $p_x(t)$ is given by \eqref{eq:pxt}.
Since $\tilde{p}_c$ is an order of magnitude smaller than $1$, we can approximate $p_x(t)$ by its leading-oder contributions,
\begin{align}\label{eq:pxApprox}
 p_x(t)\simeq p_d(t) + \sum_i\!^{'}\sin^2(J_{1i}(t))\ .
\end{align}
We follow again Ref.~\cite{Novais2013} and study an Ohmic bath ($r=0$, $D=2$). The function $J_{ij}(t)$ for this bath type has been provided in \eqref{eq:Jijspec}.
Note that $J_{ij}(t)$ decays inversely with distance outside of the light-cone.
Therefore, the second summand in \eqref{eq:pxApprox} diverges logarithmically with the code size $L$ at any non-zero time (up to constant prefactors of order $1$, we have $\sum_i\!^{'}\frac{1}{|{\bf R}_1-{\bf R}_i|^2}\sim\int_1^{L/2}\frac{1}{r^2}r\m{d}r\sim\log(L)$).
Correspondingly, the maximal QEC period vanishes in the thermodynamic limit (though it does so very slowly, see below).
For all other combinations of $D=2,3$ and $r=0,\pm\frac{1}{2}$, $J_{ij}(t)$ decays stronger than $|{\bf R}_i-{\bf R}_j|^{-1}$ outside of the light-cone (see Appendix~\ref{sec:induced}).
The maximal QEC period remains thus finite in the thermodynamic limit for all other bath types.

Setting the lattice constant of the surface code to unity and assuming a linear code size $L$, we can estimate
\begin{align}
 \sum_i\!^{'}\sin^2(J_{1i}(t)) \simeq 2\pi\int_0^{L/2}\m{d}R\,R\sin^2\left[\frac{\lambda^2}{2\pi^2v^2}\left(\theta(R-vt)\arcsin(vt/R) + \theta(vt-R)\frac{\pi}{2}\right)\right]\ .
\end{align}
Since we are interested in times where this sum is (still) sufficiently smaller than $1$, in particular each summand has to be much smaller than $1$.
Defining $m(t)=\min\lbrace L/2,vt\rbrace$, we find
\begin{align}\label{eq:mediatedError}
  \sum_i\!^{'}\sin^2(J_{1i}(t)) &\simeq 2\pi\int_{m(t)}^{L/2}\m{d}R\,\frac{1}{R}\left(\frac{\lambda^2t}{2\pi^2v}\right)^2 + 2\pi\int_0^{m(t)}\m{d}R\,R\left(\frac{\lambda^2}{4\pi v^2}\right)^2 \nn\\
&=\frac{\lambda^4t^2}{2\pi^3v^2}\log(\frac{L/2}{m(t)}) + \frac{\lambda^4}{16\pi v^4}m^2(t)\ .
\end{align}
Combining Eqs.~(\ref{eq:pdFinal}), (\ref{eq:pxApprox}), and (\ref{eq:mediatedError}) we conclude that for times $t$ which are small enough such that $p_x(t)\ll1$ we have
\begin{align}
 p_x(t) \simeq \underbrace{\frac{1}{2} - \frac{1}{2}\left[\frac{\beta\omg_c}{\pi}\sinh(\frac{\pi t}{\beta})\right]^{-2\lambda^2/\pi v^2}}_{A(t)} + \underbrace{\frac{\lambda^4t^2}{2\pi^3v^2}\log(\frac{L/2}{m(t)})}_{B(t)} + \underbrace{\frac{\lambda^4}{16\pi v^4}m^2(t)}_{C(t)}\ .
\end{align}

We can recognize three different mechanisms contributing to the single-qubit error rate $p_x(t)$. 
Summand $A(t)$ describes errors due to each qubit coupling individually to the bath. Correspondingly, this term is independent of $L$.
It is the only term that depends on temperature and the only term that contributes if the qubits do not interact via the bath.
Summand $B(t)$ describes errors due to superluminal interactions between the qubits mediated by the bath.
It diverges logarithmically with $L$ for short enough times but vanishes once all qubits are within their mutual light-cones.
Finally, summand $C(t)$ describes errors due to subluminal interactions between the qubits. Once all qubits are within their mutual light-cones, this term reaches a time-independent constant which is proportional to the number of qubits in the code.

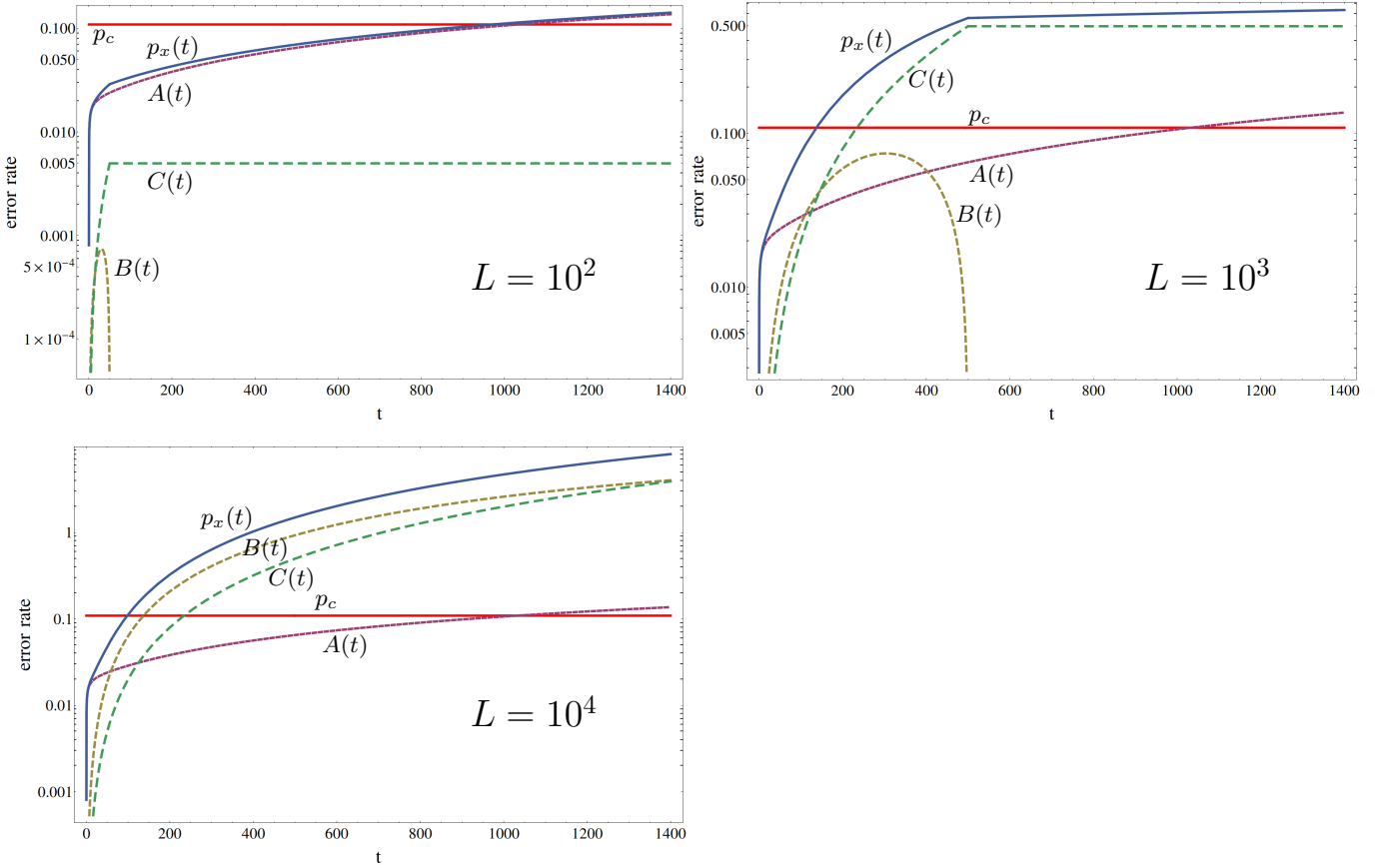
\begin{figure}[htb]
\setlength{\unitlength}{\textwidth}
\begin{picture}(0.8,0.62)
  \put(-0.11,0.32){\includegraphics[width=0.51\textwidth]{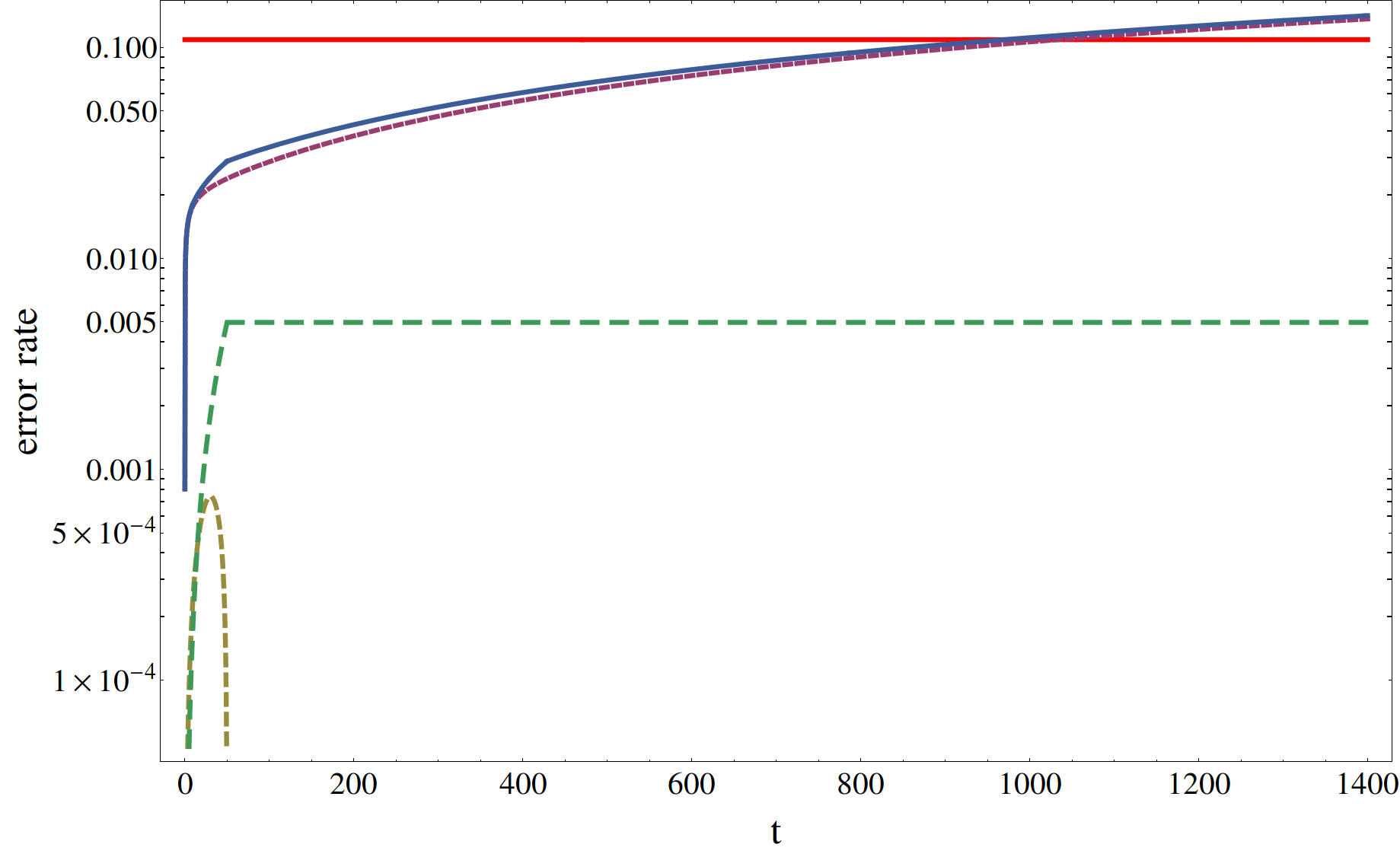}}
  \put(0.4,0.32){\includegraphics[width=0.50\textwidth]{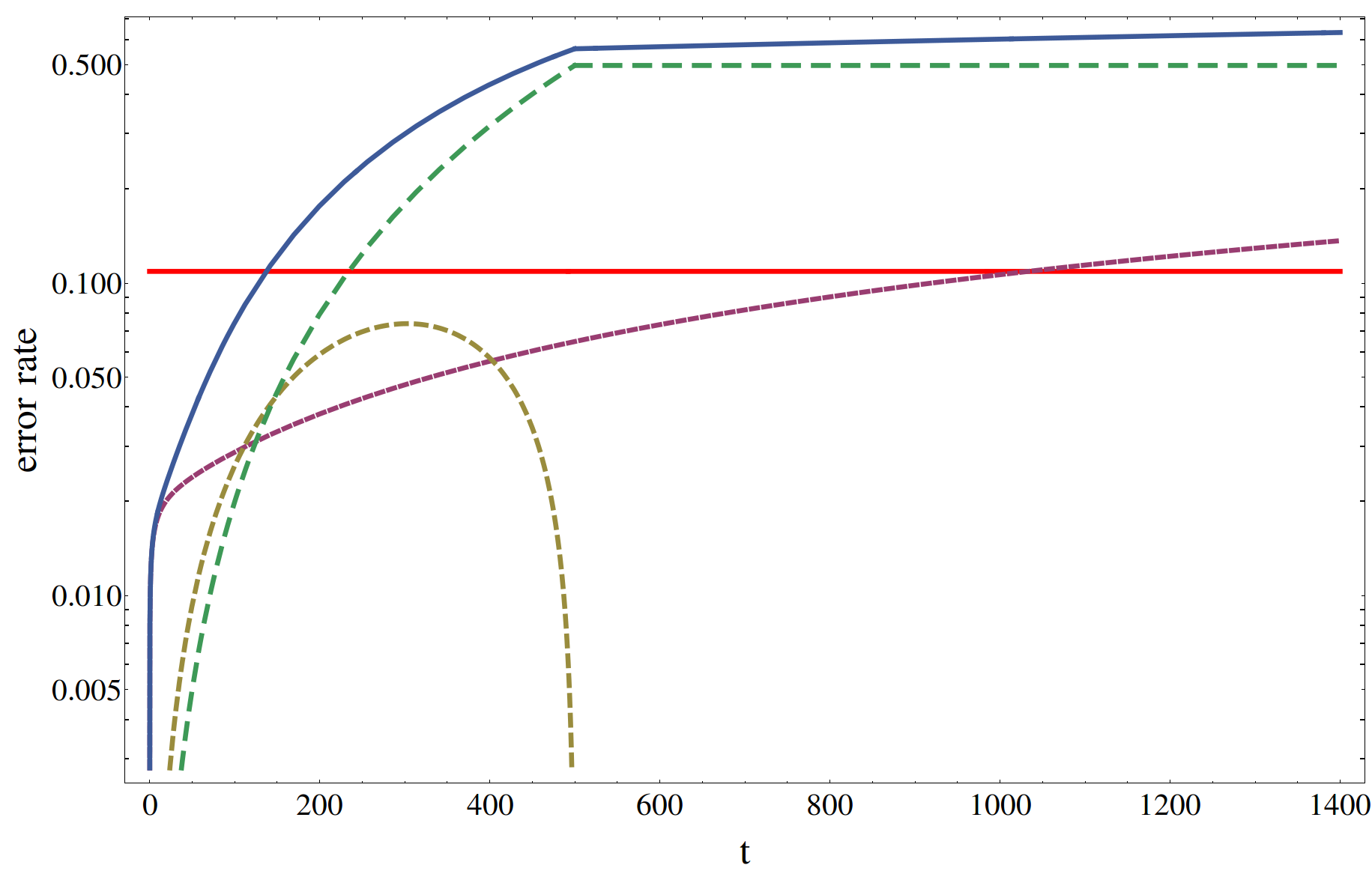}}
  \put(-0.1,-0.01){\includegraphics[width=0.50\textwidth]{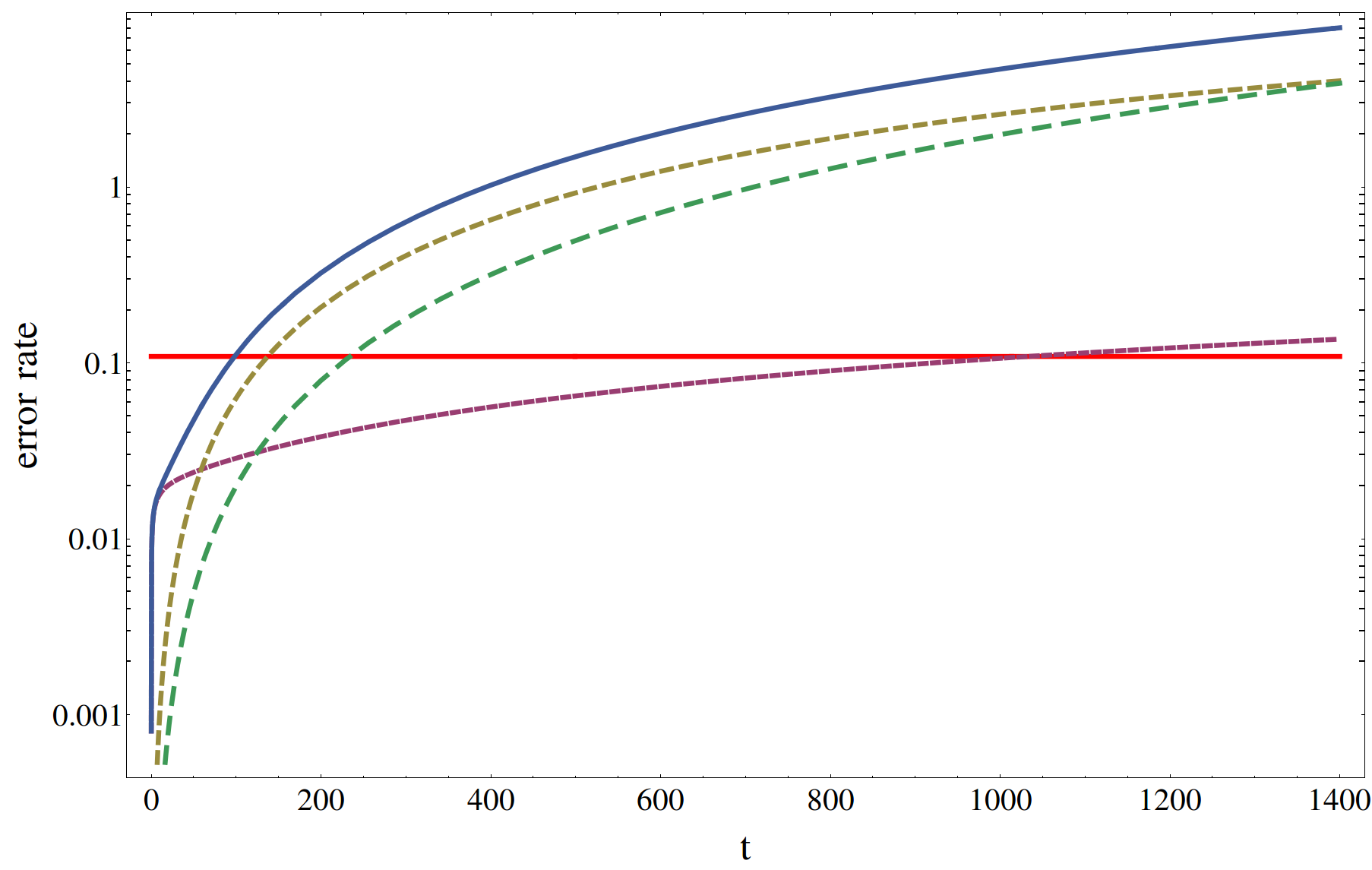}}

  \put(0.24,0.42){\Large $L=10^2$}
  \put(0.74,0.42){\Large $L=10^3$}
  \put(0.24,0.095){\Large $L=10^4$}

  \put(-0.04,0.605){$p_c$}
  \put(0.00,0.595){$p_x(t)$}
  \put(0.00,0.56){$A(t)$}
  \put(-0.025,0.43){$B(t)$}
  \put(0.00,0.495){$C(t)$}

  \put(0.61,0.545){$p_c$}
  \put(0.515,0.60){$p_x(t)$}
  \put(0.61,0.50){$A(t)$}
  \put(0.60,0.47){$B(t)$}
  \put(0.565,0.57){$C(t)$}

  \put(0.125,0.185){$p_c$}
  \put(0.04,0.245){$p_x(t)$}
  \put(0.13,0.150){$A(t)$}
  \put(0.07,0.222){$B(t)$}
  \put(0.09,0.20){$C(t)$}

\end{picture}
  \caption{The three summands $A(t)$, $B(t)$, and $C(t)$ and their sum $p_x(t)$ compared with $p_c$ for code sizes $L=10^2$, $L=10^3$, and $L=10^4$. We have used paramters $v=1$, $\lambda=0.1$, $T=0.01$, and $\omg_c=30$.
Note that the assumptions $\beta\omg_c\gg1$ and $\tau_d\gg1/\omg_c$ made during the derivation of $A(t)$ are well-satisfied.}
\label{fig:times}
\end{figure}

We have already studied the times $\tau_d$ which are necessary for summand $A(t)$ to reach critical levels ($A(\tau_d)\simeq p_c$) in Sec.~\ref{sec:uncorr}.
The only question that remains is whether $B(t)$ or $C(t)$ reach critical levels before $A(t)$ and if so, on what time-scales.
As shown in Fig.~\ref{fig:times}, each of the three summands can be the dominant force leading to the breakdown of error correction.
A higher temperature increases the weight of summand $A(t)$, while a larger code size increases the weight of summands $B(t)$ and $C(t)$.


In order to find the maximal QEC period $\tau$, we make the simplifying assumption that the breakdown is due to the dominant mechanism alone, i.e., we approximate $p_x(t)\simeq\max\lbrace A(t),B(t),C(t)\rbrace$.
Note that for times much smaller than $L/v$, we have $B(t)>C(t)$, while for times of order $L/v$ or larger, we have $B(t)<C(t)$.
For times larger than $L/2v$, $C(t)$ reaches its maximal value $\frac{\lambda^4L^2}{64\pi v^4}$. 
Therefore, for $L>8\sqrt{\pi \tilde{p}_c}\frac{v^2}{\lambda^2}$ the term $C(t)$ will reach critical values ($\tilde{p}_c$) in a time
\begin{align}\label{eq:tauSub}
 \tau_{\text{sub}}=\frac{4\sqrt{\pi \tilde{p}_c}v}{\lambda^2}\ ,
\end{align}
while otherwise it will never do so.
Elementary calculus shows that the maximal value, which $B(t)$ can achieve while still being larger than $C(t)$, is $\frac{e^{-\pi^2/4}}{64\pi}\frac{\lambda^4}{v^4}L^2$, and that $B(t)$ is monotonically increasing until it reaches this value.
Therefore, $B(t)$ reaches $\tilde{p}_c$ before $C(t)$ if and only if $L>8e^{\pi^2/8}\sqrt{\pi \tilde{p}_c}\frac{v^2}{\lambda^2}$.
The (relevant) solution to $B(\tau_{\text{super}})=\tilde{p}_c$ is given by
\begin{align}\label{eq:tauSuper}
 \tau_{\text{super}} = 2\pi\sqrt{\pi \tilde{p}_c}\frac{v}{\lambda^2} \left|W_{-1}(-16\pi^3\tilde{p}_cv^4/\lambda^4L^2)\right|^{-1/2}\ ,
\end{align}
where $W_{-1}$ is the lower branch of the Lambert $W$ function \cite{Lambert}.
For $z\ra0^{-}$, we have $W_{-1}(z)\simeq\log|z|$, showing that the available QEC time vanishes in the thermodynamic limit $L\ra\infty$ like $\tau\sim1/\sqrt{\log(L)}$, that is, very slowly.

\subsection{Summary of results}\label{sec:summary}

Let us summarize our results for a 2D Ohmic bath. There are three different mechanisms that contribute to the error rate on each qubit and hence put limits on the maximal QEC period $\tau$: 
the individual coupling of each qubit to the bath, superluminal interactions between the qubits mediated by the bath as well as subluminal ones.

The direct interaction of each qubit with the bath puts an upper bound $\tau_d$ on the maximal time for which error correction can suceed.
This time is given by \eqref{eq:tauUncorr}, for which we find a high- and a low-temperature value 
\begin{align}
 \tau_d &= \frac{1}{\pi T}\text{arcsinh}\left[\frac{\pi T}{\omg_c}(1-2\tilde{p}_c)^{-\pi v^2/2\lambda^2}\right] \nn\\
&\simeq
\begin{cases}
  \frac{1}{\omg_c}\exp(cv^2/\lambda^2) &\text{if}\quad T<\frac{\omg_c}{\pi}\exp(-cv^2/\lambda^2) \\
  \quad\\
  \frac{1}{\pi T}\left(cv^2/\lambda^2 - \log\left[\frac{\omg_c}{2\pi T}\right]\right) &\text{if}\quad T>\frac{\omg_c}{\pi}\exp(-cv^2/\lambda^2)\ .
 \end{cases}
\end{align}
Here, $c=\frac{\pi}{2}\log\frac{1}{1-2\tilde{p}_c}$. Assuming $\tilde{p}_c\simeq p_c$, we find $c\simeq0.4$.

The interaction between the qubits mediated by the bath is a further source of errors, both due to subluminal and superluminal interactions.
Errors due to mediated interactions can only reach critical values if the linear code size $L$ is large enough; if $L<8\sqrt{\pi \tilde{p}_c}\frac{v^2}{\lambda^2}$, neither the error strength due to sub- nor due to super-luminal interactions will ever reach $\tilde{p}_c$.
For $8\sqrt{\pi \tilde{p}_c}\frac{v^2}{\lambda^2}<L<8e^{\pi^2/8}\sqrt{\pi \tilde{p}_c}\frac{v^2}{\lambda^2}$, errors due to sub-luminal interaction reach a critical strength in a time $\tau_{\text{sub}}\sim v/\lambda^2$.
If errors due to superluminal interactions also reach criticality, they will do so on times larger than $\tau_{\text{sub}}$ for these values of $L$.
Finally, if $L>8e^{\pi^2/8}\sqrt{\pi \tilde{p}_c}\frac{v^2}{\lambda^2}$, superluminally meadiated errors reach criticality before subluminal ones, and they do so in a time $\tau_{\text{super}}\sim v/\lambda^2\sqrt{\log L}$.
This time vanishes very slowly in the thermodynamic limit. These results are summarized in the following table (assuming $\tilde{p}_c\simeq p_c$).

\begin{center}
    \begin{tabular}{ | l | l | l | }
    \hline
    Code size & Breakdown in a time & Dominant mechanism  \\ \hline\hline
    $L<4.7\frac{v^2}{\lambda^2}$ & $\tau_d$ & direct bath coupling \\ \hline
    $4.7\frac{v^2}{\lambda^2}<L<16.1\frac{v^2}{\lambda^2}$ & $\min\lbrace\tau_d,\tau_{\text{sub}}\rbrace$ & direct bath coupling or subluminal interactions \\ \hline
    $L>16.1\frac{v^2}{\lambda^2}$ & $\min\lbrace\tau_d,\tau_{\text{super}}\rbrace$ & direct bath coupling or superluminal interactions \\ \hline
    \end{tabular}
\end{center}


\section{Conclusions}\label{sec:conclusions}

Quantum information is fragile and can only be maintained if the accumulation of entropy in the information-bearing degrees of freedom of a storage device can be suppressed -- either by preventing entropy from entering or by removing it at a sufficient pace.     
Any possible measure to achive this can only succeed for certain classes of system-environment couplings.
Correspondingly, a proposal that promises stability of quantum information is only as valuable as the error source against which it protects is realistic.

In this work, we have investigated how long the surface code is able to protect a quantum state against noise emerging from a physically relevant type of environment -- a bath of freely propagating bosonic modes.
We have seen that there are two very distince kinds of error mechanisms: the individual decoherence of each qubit, and induced interactions between the code qubits.
Both mechanisms lead to spatial and temporal correlations between the errors happening in the code.
However, we have shown that a tendency of errors to cluster without a tendency to form string-like configurations does not strongly derogate the correctability of the surface code -- even when these correlations are ignored during error correction.

We have managed to express the time before the error rates in the code reach critical values in terms of code size ($L$), accidental coupling strength ($\lambda$), mode velocity ($v$), and bath temperature ($T$) across a wide range of different parameter regimes.
Two further parameters that determine the physical character of the qubits' decoherence mechanism are the spatial dimension of the medium in which the modes propagate ($D$) and the nature of the coupling to the bath ($r$).
We have focused our discussion on the specific combination ($D=2$, $r=0$) investigated in Ref.~\cite{Novais2013}, which corresponds to an Ohmic bath. 
This combination is of particular interest since it is the only one for which the maximal QEC time vanishes (very slowly) in the thermodynamic limit.
For all other combinations of $D=2,3$ and $r=0,\pm\frac{1}{2}$, this time remains finite.

Following Refs.~\cite{Novais2013,Jouzdani2013}, we have made several simplifying assumptions to make the actual problem analytically tractable. 
These are: a trivial Hamiltonian for the qubits; undamped and non-interacting bath modes; no residual bath correlations between different QEC periods; one type of errors only (bit-flips); immediate and flawless syndrome measurement and error correction 
(including no time cost for efficient classical computations).
Relaxing these assumptions opens a wide field of additional challenges. 
For instance, fully fault-tolerant syndrome extraction and error correction are discussed in Ref.~\cite{Wang2011,Duclos2014}.
A finite probability of syndrome measurement failure will lead to a lower value of $\tilde{p}_c$ and hence necessitate shorter QEC periods.
Moreover, we have in this work been concerned exclusively with spatial and temporal correlations between errors in the surface code.
If there are non-commuting error types on the same qubit (bit- and phase-flips), a further type of correlation in the noise emerges, namely correlations between different error types on the same qubit.
Such correlations are present in the often-used error model of depolarizing noise. How they can be taken into account during error correction is studied in Refs.~\cite{Duclos2009,Wootton2012,Hutter2013,Fowler2013}.
Finally, adding an energy splitting $-\frac{\Delta}{2}\tilde{\sum}_i\pz_i$ for the code qubits would transform the problem into a many-spin generalization of the well-studied spin-boson problem.
For a single spin-qubit coupled to an Ohmic bath, the spin-boson problem has been solved within the Born approximation in Ref.~\cite{DiVincenzo2005}. 
However, the generalization of this problem to the many-qubit case may well be analytically intractable \cite{Terhal2005}.


\section{Acknowledgements}

We would like to thank J.~R. Wootton, P. Jouzdani, B.~M. Terhal, and A.~G. Fowler for helpful discussions. This work was supported by the Swiss NF, NCCR QSIT, and IARPA.

\appendix

\section{Different bath types}\label{app:different}

\subsection{Induced interactions}\label{sec:induced}

\subsubsection{Linear dispersion}

The bath-induced pairwise interaction between code qubits is described by the function
\begin{align}
 J_{ij}(t) = 2\lambda^2\int\m{d}{\bf k}\,e^{-v|{\bf k}|/\omg_c}\frac{|{\bf k}|^{2r}}{\omg\kk^2}\cos\left({\bf k}({\bf R}_i-{\bf R}_j)\right)\left(\sin(\omg\kk t)-\omg\kk t\right)\,,
\end{align}
where we have introduced a cut-off factor $e^{-v|{\bf k}|/\omg_c}$ into the expression given in \eqref{eq:Jij}. 
The cut-off factor is only necessary in the case (3D, $r=\frac{1}{2}$), while in all other cases we can let $\omg_c\ra\infty$.
In 2D, the functions $J_{ij}(t)$ can be calculated as described in Ref.~\cite[Appendix C]{Jouzdani2013}.
With $\omg\kk=v|{\bf k}|$ and $R:=|{\bf R}_i-{\bf R}_j|$, the results are
\begin{align}
 J_{ij}(t) =
\begin{cases}
 \frac{\lambda^2}{2\pi^2v^2}\theta(vt-R)\left(\sqrt{v^2t^2-R^2}-vt\log(\frac{vt+\sqrt{v^2t^2-R^2}}{R})\right) &\text{for }r=-\frac{1}{2} \\
 \frac{\lambda^2}{2\pi^2v^2}\left(\theta(R-vt)\arcsin(vt/R) + \theta(vt-R)\frac{\pi}{2}\right) &\text{for }r=0 \\
 \frac{\lambda^2}{2\pi^2v^2}\frac{\theta(vt-R)}{\sqrt{v^2t^2-R^2}} &\text{for }r=\frac{1}{2}\,,
\end{cases}
\end{align}
while in 3D, we find
\begin{align}
 J_{ij}(t) =
\begin{cases}
 -\frac{\lambda^2}{2\pi Rv^2}(vt-R)\theta(vt-R) &\text{for }r=-\frac{1}{2} \\
 \frac{\lambda^2}{2\pi^2Rv^2}\left(\log\left|\frac{R+vt}{R-vt}\right|-\frac{2vt}{R}\right) &\text{for }r=0 \\
 \frac{2\lambda^2}{\pi^2R^4v\omg_c}\frac{2R^2-v^2t^2}{(R^2-v^2t^2)^2}v^3t^3 &\text{for }r=\frac{1}{2}\,.
\end{cases}
\end{align}

Note that in three cases the interaction vanishes exactly outside of the light-cone.
The combination (2D, $r=0$) considered in the main text shows the longest-range superluminal interactions.
It is the only one for which the sum $\sum_i\!^{'}\sin^2(J_{1i}(t))$ in \eqref{eq:pxApprox} diverges for any non-zero time in the thermodynamic limit.
Correspondingly, it is the only one for which the maxmial QEC period (theoretically) vanishes in this limit.

\subsubsection{Ordered ferromagnet: parabolic dispersion}\label{sec:parabolic}

Recently, the idea of performing entangling gates between qubits by coupling them to an ordered Heisenberg ferromagnet has attracted interest \cite{Trifunovic2013}.
An ordered Heisenberg ferromagnet can be seen as a 3D magnon bath ($\omg\kk=D{\bf k}^2$). If we couple to a spin component which is orthogonal to the ordering, we obtain a coupling of type $r=0$.
The ferrogmanet thus acts as a sub-Ohmic bath ($s=\frac{1}{2}$).
Then,
\begin{align}
 J_{ij}(t) = \frac{\lambda^2}{4\pi^2D^2R}\left[-2\pi Dt -\pi(R^2-2Dt)C\left(\frac{R}{\sqrt{2\pi Dt}}\right) +\pi(R^2+2Dt)S\left(\frac{R}{\sqrt{2\pi Dt}}\right) + \sqrt{2\pi Dt}R(\cos(\frac{R^2}{4Dt})+\sin(\frac{R^2}{4Dt}))\right]\,,
\end{align}
where $C(x)=\int_0^x\cos(t^2)\m{d}t$ and $S(x)=\int_0^x\sin(t^2)\m{d}t$ are the Fresnel-integrals. 
For times such that $R\ll\sqrt{Dt}$, the first summand in the bracket dominates and we find
\begin{align}
 J_{ij}(t) = -\frac{\lambda^2t}{2\pi DR}\,.
\end{align}

\subsection{Decoherence}\label{sec:decoherence}

We have shown in Sec.~\ref{sec:uncorr} that the probability of an error due to the coupling of a qubit to the bath is given by
\begin{align}
 p_d(t) = \frac{1}{2}\left(1-\exp\left\lbrace-2\Lambda(t)\right\rbrace\right)\,,
\end{align}
where the function $\Lambda(t)$ is given in \eqref{eq:integral}.
It depends only on the spectral function $J(\omg)=\alpha\omg^s\omg_0^{1-s}e^{-\omg/\omg_c}$ of the bath and its temperature.
The cases $s=0,1,2,3$ are relevant for the kinds of couplings to a bath with linear dispersion in 2D or 3D considered in the main part of this work.
The case $s=\frac{1}{2}$ is relevant for an ordered Heisenberg ferromagnet (see previous subsection).

For those values of $s$, we find
\begin{align}\label{eq:Lambdas}
 \Lambda(t) = 
\begin{cases}
\alpha\pi\omega_0t & \m{if}\quad s=0 \quad\m{and}\quad\beta\ra\infty \\
2\alpha\sqrt{2\pi\omg_0t}+2\alpha\sqrt{\frac{\beta\omg_0}{\pi}}\left(4\pi\m{Re}\,\zeta(-\frac{1}{2},1+\frac{it}{\beta})+\zeta(\frac{3}{2})\right) & \m{if}\quad s=\frac{1}{2} \quad\m{and}\quad \beta\omg_c\gg1 \\
\alpha\log(1+\omg_c^2t^2)+2\alpha\log\left(\frac{\beta}{\pi t}\sinh(\frac{\pi t}{\beta})\right) & \m{if}\quad s=1 \quad\m{and}\quad \beta\omg_c\gg1 \\
\frac{\alpha}{\omg_0}\left(-\frac{2\omg_c^2t^2}{1+\omg_c^2t^2}-\frac{4}{\beta}\psi(\frac{1}{\beta\omg_c})+\frac{4}{\beta}\m{Re}\,\psi(\frac{1+i\omg_ct}{\beta\omg_c})\right) & \m{if}\quad s=2 \\
\frac{\alpha}{\omg_0^2}\left(-\frac{2\omg_c^4t^2(3+\omg_c^2t^2)}{(1+\omg_c^2t^2)^2}+\frac{4}{\beta^2}\psi'(\frac{1}{\beta\omg_c})-\frac{4}{\beta^2}\m{Re}\,\psi'(\frac{1+i\omg_ct}{\beta\omg_c})\right) & \m{if}\quad s=3\ .
\end{cases}
\end{align}
Here, $\zeta(-\frac{1}{2},z)$ denotes a Hurwitz zeta function and $\psi(z)$ is the digamma function.
The case $s=0$ requires at finite temperature an infrared cut-off for convergence.
The result for $s=1$ has been derived in \eqref{eq:Lambda}, the result for $s=\frac{1}{2}$ can be derived in a very analogous way.
Note that a well-defined $\omg_c\ra\infty$ limit exists only for sub-Ohmic baths. 

\begin{figure}
\setlength{\unitlength}{0.8\textwidth}
  \begin{picture}(1.0,0.5)
	\put(0.0,0.0){\includegraphics[width=0.8\textwidth]{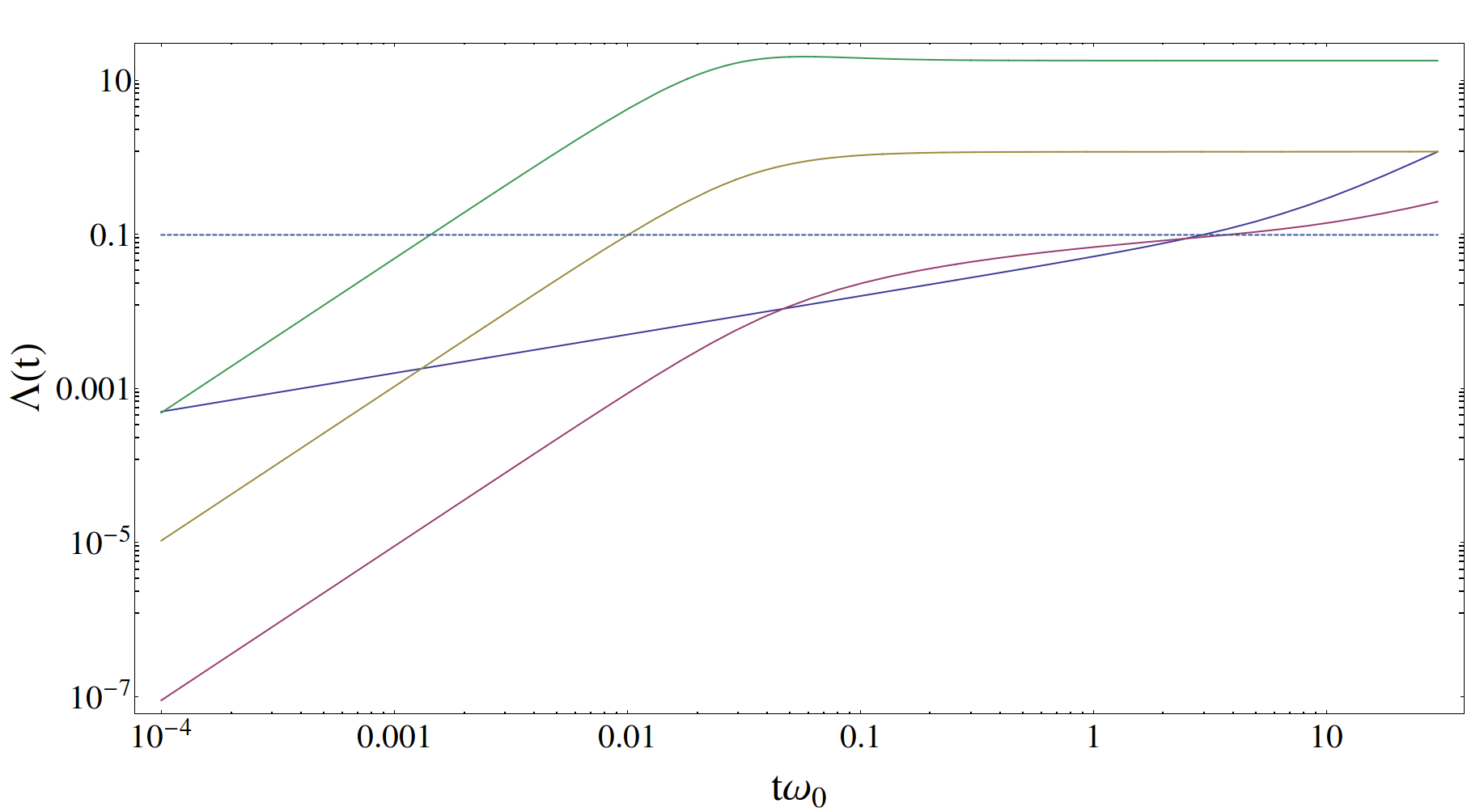}}
 	\put(0.39,0.34){$s=\frac{1}{2}$}
 	\put(0.39,0.25){$s=1$}
 	\put(0.39,0.405){$s=2$}
 	\put(0.39,0.49){$s=3$}
  \end{picture}
 \caption{\label{fig:lambdas} 
  The bold lines show the functions $\Lambda(t)$ given in \eqref{eq:Lambdas}.
  We have used parameters $\alpha=0.01$, $\omg_c/\omg_0=30$, and $\beta\omg_0=10$.
  The dashed line shows the critical value $\frac{1}{2}\log\frac{1}{1-2p_c} \simeq 0.123$, when error correction breaks down in the uncorrelated case.}
\end{figure}
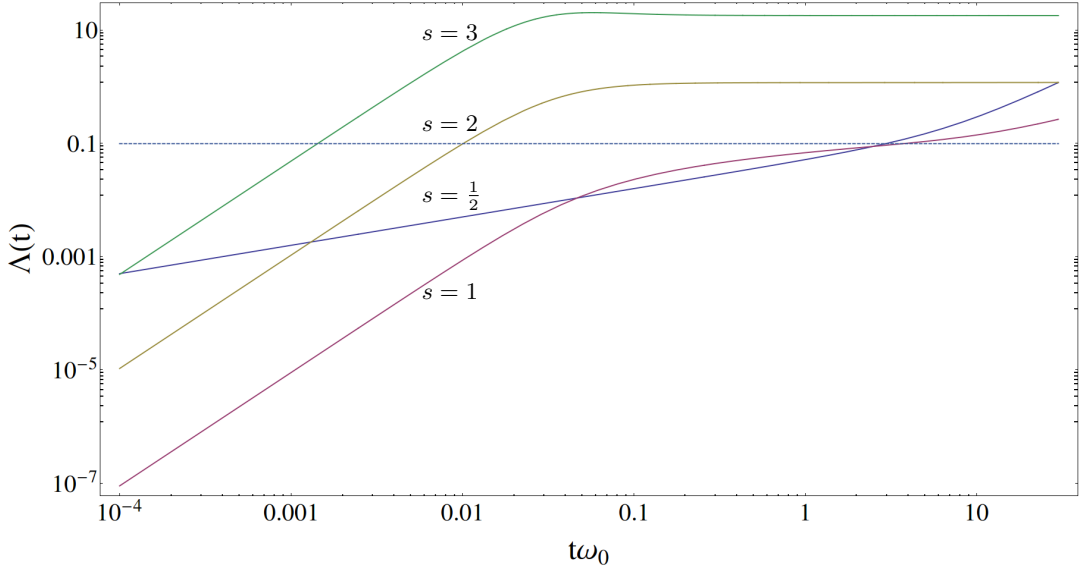

The expressions in \eqref{eq:Lambdas} (for $s>0$) are displayed for a specific set of parameters $\alpha$, $\omg_0$, $\omg_c$, and $\beta$ in Fig.~\ref{fig:lambdas} 
and compared to the critical value of $\Lambda(t)$ in the case of uncorrelated errors given in \eqref{eq:tauEq}.
We see that for super-Ohmic baths this critical value is reached distinctively earlier than for Ohmic and sub-Ohmic baths.

\section{An algorithm that is able to take correlations between errors on nearest neighbors into account}\label{app:algo}

We assume again a single-qubit error rate $p_1$ and a rate of two-qubit errors on nearest neighbors $p_2$.
To an edge connecting anyons $i$ and $j$, we want to assign a weight $-\log(p_{ij})$, where $p_{ij}$ is the sum of the probabilities of all error chains connecting anyons $i$ and $j$.
The minimal-weight error chain is then the most likely one.
Taking the negative logarithm ensures that the weights are additive for independent error chains.
More precisely, we will not consider the absolute probabilities but the probabilities relative to no errors happening.
This leads to a constant shift of all weights, which is irrelevant since the number of edges involved in each matching is identical.

Using the Manhattan distance of the anyons as the weight, as we did for the algorithm that ignores correlations between errors, 
corresponds to approximating $p_{ij}$ by the probability of the most likely single-qubit error path connecting anyons $i$ and $j$, without taking the degeneracy of this probability into account.
While calculating $p_{ij}$ exactly is unfeasible, the algorithm presented here is based on a better approximation of $p_{ij}$, which, in particular, takes the possibility of two-qubit errors into account.
We will restrict to those error chains which probabilistically dominate for either $p_1\gg p_2$ or $p_1\ll p_2$.

Assume that anyons $i$ and $j$ have horizontal distance $a$ and vertical distance $b$, or \emph{vice versa}, with $a\leq b$.
Below, we list all contributions to $p_{ij}$ which we consider. The probability-independent prefactors are the number of possible paths of the respective type.
We denote with $m_1$ the number of one-qubit events and with $m_2$ the number of two-qubit events in an error path.
We consider all error paths contributing to $p_{ij}$ for which $m_1\leq1$ or $m_2\leq1$, and which are such that there is no error path with error numbers $m_1'$ and $m_2'$ connecting anyons $i$ and $j$ such that $m_1'\leq m_1$, $m_2'\leq m_2$, and $m_1'+m_2'<m_1+m_2$.

\begin{center}
    \begin{tabular}{ | l | l | l | }
    \hline
    Condition & Type & Contribution to $p_{ij}$  \\ \hline\hline
    \quad & $m_2=0$ & $\binom{a+b}{a}\left(\frac{p_1}{1-p_1}\right)^{a+b}$ \\ \hline
    $a + b \equiv 0\,\,\, (\m{mod}\,\,2)$ & $m_1=0$ & $\binom{b}{(b-a)/2} \left(\frac{p_2}{1-p_2}\right)^b$ \\ \hline
    $a + b \equiv 1\,\,\, (\m{mod}\,\,2) \land a+b\geq3$ & $m_1=1 \land m_2\geq1$ & $\frac{a+b+1}{2}\binom{b}{(b-a-1)/2}\frac{p_1}{1-p_1}\left(\frac{p_2}{1-p_2}\right)^{b-1}$ \\ \hline
    $a\geq1 \land a+b\geq4$ & $m_1\geq2 \land m_2=1$ & $(a+b-1)\binom{a+b-2}{a-1}\left(\frac{p_1}{1-p_1}\right)^{a+b-2}\frac{p_2}{1-p_2}$ \\ \hline
    \end{tabular}
\end{center}

\section{Exact evolution of two-qubit density matrix}\label{app:coupling}

Let us assume that only two qubits, $i$ and $j$, couple to the bath and let us study their joint evolution, which according to Eqs.~(\ref{eq:Phid}) and (\ref{eq:U}), is given by
\begin{align}
 \rho_{ij}(t) = \exp\left\lbrace-iJ_{ij}(t)\px_i\otimes\px_j\right\rbrace \EE_t(\rho_{ij}) \exp\left\lbrace+iJ_{ij}(t)\px_i\otimes\px_j\right\rbrace\ .
\end{align}
where
\begin{align}\label{eq:rhoij}
 \EE_t(\rho_q) &= \tr_B\left\lbrace e^{\px_i\otimes X_i(t)}e^{\px_j\otimes X_j(t)}(\rho_{ij}\otimes\rho_B)e^{-\px_i\otimes X_i(t)}e^{-\px_j\otimes X_j(t)} \right\rbrace \nn\\
&= 
\rho_{ij}\times\left\langle\cosh^2(X_i(t))\cosh^2(X_j(t))\right\rangle
-\px_i\rho_{ij}\px_i\times\left\langle\sinh^2(X_i(t))\cosh^2(X_j(t))\right\rangle \nn\\&\quad
-\px_j\rho_{ij}\px_j\times\left\langle\cosh^2(X_i(t))\sinh^2(X_j(t))\right\rangle
+\px_i\px_j\rho_{ij}\px_i\px_j\times\left\langle\sinh^2(X_i(t))\sinh^2(X_j(t))\right\rangle \nn\\&\quad
+\left(\px_i\px_j\rho_{ij}+\rho_{ij}\px_i\px_j-\px_i\rho_{ij}\px_j-\px_j\rho_{ij}\px_i\right) \times \left\langle\cosh(X_i(t))\sinh(X_i(t))\cosh(X_j(t))\sinh(X_j(t))\right\rangle\ .
\end{align}

Our goal is to express all appearing expectation values in terms of the correlators
\begin{align}
 C_{ij}(t) = \langle X_i(t)X_j(t)\rangle = -\frac{\lambda^2}{N}\sum\kk|{\bf k}|^{2r}\cos\left({\bf k}({\bf R}_i-{\bf R}_j)\right)\coth(\beta\omg\kk/2)\frac{\sin^2(\omg\kk t/2)}{(\omg\kk/2)^2}\ .
\end{align}
The fuctions $\Lambda(t)=-C_{ii}(t)$ are discussed in detail in Sec.~\ref{sec:uncorr}.

Recall that  $\sinh^2(x)=\frac{1}{2}(\cosh(2x)-1)$ and $\cosh^2(x)=\frac{1}{2}(\cosh(2x)+1)$.
The first four expectation values (those corresponding to diagonal terms) can thus be reduced to $\langle\cosh(2X_i(t))\rangle$ and $\langle\cosh(2X_i(t))\cosh(2X_j(t))\rangle$.
We already know that 
\begin{align}
 \langle\cosh(2X_i(t))\rangle = 2\langle\sinh^2(X_i(t))\rangle + 1 = \exp\lbrace-2\Lambda(t)\rbrace
\end{align}
(see \eqref{eq:pd}).
Let us thus calculate
 \begin{align}
  \langle\cosh(2X_i(t))\cosh(2X_j(t))\rangle &= \sum_{m,n=0}^\infty\frac{2^{2m+2n}}{(2m)!(2n)!}\langle X_i(t)^{2m}X_j(t)^{2n}\rangle  \nn\\
&= \sum_{m,n=0}^\infty\sum_{k}^{\min(m,n)}\frac{2^{2m+2n}}{(2m)!(2n)!}\binom{2m}{2k}(2k)!\binom{2n}{2k}\langle X_i(t)^{2m-2k}\rangle\langle X_i(t)X_j(t)\rangle^{2k}\langle X_j(t)^{2n-2k}\rangle  \nn\\
&= \sum_{m,n=0}^\infty\sum_{k}^{\min(m,n)}(-1)^{m+n}\frac{2^{2m+2n}}{(2m)!(2n)!}\binom{2m}{2k}(2k)!\binom{2n}{2k}\frac{(2m-2k)!}{2^{m-k}(m-k)!}\frac{(2n-2k)!}{2^{n-k}(n-k)!} \nn\\&\quad\times \Lambda(t)^{m+n-2k}C_{ij}(t)^{2k} \nn\\
&= \sum_{m,n=0}^\infty\sum_{k}^{\min(m,n)} \frac{(-2)^{m+n+2k}\Lambda(t)^{m+n-2k}C_{ij}(t)^{2k}}{(m-k)!(n-k)!(2k)!} \ .
 \end{align}
To simplify this expression, we define $u:=m-k$ and $v:=n-k$. Then, 
 \begin{align}
  \langle\cosh(2X_i(t))\cosh(2X_j(t))\rangle &= 
  \sum_{u,v,k=0}^\infty  \frac{(-2)^{u+v+4k}\Lambda(t)^{u+v}C_{ij}(t)^{2k}}{u!v!(2k)!} \nn\\
 &= e^{-4\Lambda(t)} \cosh(4C_{ij}(t))\ .
\end{align}
We conclude that 
\begin{align}
 \left\langle\cosh^2(X_i(t))\cosh^2(X_j(t))\right\rangle &= \frac{1}{4}+\frac{1}{2}e^{-2\Lambda(t)}+\frac{1}{4}e^{-4\Lambda(t)}\cosh(4C_{ij}(t))\ , \nn\\
 \left\langle\sinh^2(X_i(t))\cosh^2(X_j(t))\right\rangle &= \left\langle\cosh^2(X_i(t))\sinh^2(X_j(t))\right\rangle = \frac{1}{4}e^{-4\Lambda(t)}\cosh(4C_{ij}(t))-\frac{1}{4}\ ,\text{ and} \nn\\
 \left\langle\cosh^2(X_i(t))\sinh^2(X_j(t))\right\rangle &= \frac{1}{4}-\frac{1}{2}e^{-2\Lambda(t)}+\frac{1}{4}e^{-4\Lambda(t)}\cosh(4C_{ij}(t))\ .
\end{align}
Let us now also calculate the remaining expectation value in \eqref{eq:rhoij}. We find
\begin{align}
 &\left\langle\cosh(X_i(t))\sinh(X_i(t))\cosh(X_j(t))\sinh(X_j(t))\right\rangle \nn\\
&\quad= \frac{1}{4} \left\langle\sinh(2X_i(t))\sinh(2X_j(t))\right\rangle \nn\\
&\quad= \frac{1}{4} \sum_{m,n=0}^\infty \frac{2^{2m+1}}{(2m+1)!}\frac{2^{2n+1}}{(2n+1)!} \left\langle X_i(t)^{2m+1}X_j(t)^{2n+1}\right\rangle \nn\\
&\quad= \frac{1}{4} \sum_{m,n=0}^\infty\sum_{k=0}^{\min(m,n)} \frac{2^{2m+1}}{(2m+1)!}\frac{2^{2n+1}}{(2n+1)!} \binom{2m+1}{2k+1}(2k+1)!\binom{2n+1}{2k+1}
    \left\langle X_i(t)^{2m-2k}\right\rangle\left\langle X_i(t)X_l(t)\right\rangle^{2k+1}\left\langle X_j(t)^{2n-2k}\right\rangle \nn\\
&\quad= \frac{1}{4} \sum_{m,n=0}^\infty\sum_{k=0}^{\min(m,n)} (-1)^{m+n} \frac{2^{2m+1}}{(2m+1)!}\frac{2^{2n+1}}{(2n+1)!} \binom{2m+1}{2k+1}(2k+1)!\binom{2n+1}{2k+1}
    \frac{(2m-2k)!}{2^{m-k}(m-k)!}\frac{(2n-2k)!}{2^{n-k}(n-k)!} \nn\\&\qquad\times \Lambda(t)^{m+n-2k}C_{ij}(t)^{2k+1} \nn\\
&\quad= \sum_{m,n=0}^\infty\sum_{k=0}^{\min(m,n)}\frac{(-2)^{m+n+2k}}{(m-k)!(n-k)!(2k+1)!} \Lambda(t)^{m+n-2k}C_{ij}(t)^{2k+1} \nn\\
&\quad= \sum_{u,v,k=0}^\infty\frac{(-2)^{u+v+4k}}{u!v!(2k+1)!} \Lambda(t)^{u+v}C_{ij}(t)^{2k+1} \nn\\
&\quad= \frac{1}{4}e^{-4\Lambda(t)}\sinh(4C_{ij}(t))\ .
\end{align}
Therefore,
\begin{align}
 \rho_{ij}(t) &= \exp\left\lbrace -iJ_{ij}(t)\px_i\otimes\px_j\right\rbrace \left[
\left(\frac{1}{4}+\frac{1}{2}e^{-2\Lambda(t)}+\frac{1}{4}e^{-4\Lambda(t)}\cosh(4C_{ij}(t))\right) \times \rho_{ij} \right. \nn\\&\quad
+\left(\frac{1}{4}-\frac{1}{4}e^{-4\Lambda(t)}\cosh(4C_{ij}(t))\right) \times (\px_i\rho_{ij}\px_i + \px_j\rho_{ij}\px_j) \nn\\&\quad
+\left(\frac{1}{4}-\frac{1}{2}e^{-2\Lambda(t)}+\frac{1}{4}e^{-4\Lambda(t)}\cosh(4C_{ij}(t))\right) \times \px_i\px_j\rho_{ij}\px_i\px_j \nn\\&\quad
\left. +\frac{1}{4}e^{-4\Lambda(t)}\sinh(4C_{ij}(t)) \times \left(\px_i\px_j\rho_{ij}+\rho_{ij}\px_i\px_j-\px_i\rho_{ij}\px_j-\px_j\rho_{ij}\px_i\right) \right] \exp\left\lbrace +iJ_{ij}(t)\px_i\otimes\px_j\right\rbrace\ .
\end{align}
Using now $\exp\left\lbrace \pm iJ_{ij}(t)\px_i\otimes\px_j\right\rbrace = \cos(J_{ij}(t))\pm i\sin(J_{ij}(t))\px_i\otimes\px_j$, we arrive at \eqref{eq:2qubitfinal}.


\begin{thebibliography}{99}
\bibitem{Lambert}
The Lambert $W$ function by definition satisfies $z=W(z)e^{W(z)}$. For $-e^{-1}<z<0$, there are two solutions, giving rise to two branches $W_{-1}(z)$ and $W_0(z)$ with $W_{-1}(z)<W_0(z)<0$.
The equation $a=x^2\log(\xi/x)$ with $a>0$ and $\xi>\sqrt{2ae}$ has the two solutions $x=\sqrt{2a}\left|W_k(-2a/\xi^2)\right|^{-1/2}$, with $k=-1,0$. 
Since we are interested in the smaller of the two solutions, we choose the $k=-1$ branch and obtain \eqref{eq:tauSuper}.
\end{thebibliography}
\end{document}